\documentclass[12pt, draftclsnofoot, onecolumn]{IEEEtran}

\usepackage{amsmath}
\usepackage{amssymb}
\usepackage{graphicx}
\usepackage{tikz}
\usepackage{tkz-tab}
\usetikzlibrary{patterns}  
\usepackage{pgfplots}
\usepackage{booktabs}
\usepackage{floatflt}
\usepackage{enumerate}
\usepackage{psfrag}	
\usepackage{array}
\usepackage{multirow,hhline}
\usepackage{exscale}
\usepackage{color}
\usepackage{colortbl}
\usepackage{epsfig}
\usepackage{cite}
\usepackage{amsthm}
\usepackage{array}
\usepackage{subfigure}

\usepackage[font=small]{caption}
\usepackage[latin1]{inputenc} 
\usepackage[T1]{fontenc} 
\usepackage{rotating}
\usepackage{pbox}

\usepackage{algorithm}
\usepackage{algpseudocode}
\usepackage{pifont}


\newcommand\given[1][]{\:#1\vert\:}

\newcommand{\indep}{\rotatebox[origin=c]{90}{$\models$}}

\newtheorem{remark}{Remark}

\usetikzlibrary{shapes,snakes}
\usetikzlibrary{calc}
\usetikzlibrary{patterns}
\usetikzlibrary{decorations.pathmorphing} 

\newcommand{\iPhoneHalfDuplex}[3]{
\coordinate (a) at (#1,#2);
\draw [line width=0.25pt,rounded corners=(#3)*1mm,fill=white,scale=(#3)] (a)--($(a)+(0.67,0)$)--($(a)+(0.67,1.381)$)--($(a)+(0,1.381)$)--cycle;
\draw [color=gray,line width=0.25pt,rounded corners=(#3)*0.8mm,fill=white,scale=(#3)] ($(a)+(0.015,0.015)$)--($(a)+(0.655,0.015)$)--($(a)+(0.655,1.366)$)--($(a)+(0.015,1.366)$)--cycle;
\draw [line width=0.25pt,rounded corners=(#3)*0.04mm,scale=(#3)] ($(a)+(0.2875,1.266)$)--($(a)+(0.3825,1.266)$)--($(a)+(0.3825,1.281)$)--($(a)+(0.2875,1.281)$)--cycle;
\draw[line width=0.25pt,scale=#3] ($(a)+(0.335,0.09)$) circle (0.055cm);
\draw[line width=0.25pt,scale=#3] ($(a)+(0.335,0.09)$) circle (0.044cm);
\draw[line width=0.25pt,scale=#3] ($(a)+(0.2275,1.2735)$) circle (0.015cm);
\draw[line width=0.25pt,scale=#3] ($(a)+(0.335,1.32)$) circle (0.01cm);
\draw [fill={rgb:black,1;white,4},line width=0.25pt,scale=(#3)] ($(a)+(0.042475,0.170195)$)--($(a)+(0.042475,0.170195)+(0.58505,0.0)$)--($(a)+(0.042475,0.170195)+(0.58505,1.04061)$)--($(a)+(0.042475,0.170195)+(0.0,1.04061)$)--cycle;
}

\newcommand{\iPhoneFullDuplex}[3]{
\coordinate (a) at (#1,#2);
\draw [line width=0.25pt,rounded corners=(#3)*1mm,fill=white,scale=(#3)] (a)--($(a)+(0.67,0)$)--($(a)+(0.67,1.381)$)--($(a)+(0,1.381)$)--cycle;
\draw [color=red,line width=0.25pt,rounded corners=(#3)*0.8mm,fill=white,scale=(#3)] ($(a)+(0.015,0.015)$)--($(a)+(0.655,0.015)$)--($(a)+(0.655,1.366)$)--($(a)+(0.015,1.366)$)--cycle;
\draw [line width=0.25pt,rounded corners=(#3)*0.04mm,scale=(#3)] ($(a)+(0.2875,1.266)$)--($(a)+(0.3825,1.266)$)--($(a)+(0.3825,1.281)$)--($(a)+(0.2875,1.281)$)--cycle;
\draw[line width=0.25pt,scale=#3] ($(a)+(0.335,0.09)$) circle (0.055cm);
\draw[line width=0.25pt,scale=#3] ($(a)+(0.335,0.09)$) circle (0.044cm);
\draw[line width=0.25pt,scale=#3] ($(a)+(0.2275,1.2735)$) circle (0.015cm);
\draw[line width=0.25pt,scale=#3] ($(a)+(0.335,1.32)$) circle (0.01cm);
\draw [fill={rgb:black,1;red,4},line width=0.25pt,scale=(#3)] ($(a)+(0.042475,0.170195)$)--($(a)+(0.042475,0.170195)+(0.58505,0.0)$)--($(a)+(0.042475,0.170195)+(0.58505,1.04061)$)--($(a)+(0.042475,0.170195)+(0.0,1.04061)$)--cycle;
}

\newcommand{\basestationHalfDuplex}[3]{
\coordinate (a) at (#1,#2);
\draw[line width=(#3)*1.5pt,scale=#3] ($(a)+(0, -1.08)$) -- ($(a)+(0, 1)$);
\draw[line width=(#3)*1.5pt,scale=#3] ($(a)+(0,1)$) .. controls ($(a)+(-0.25,-0.5)$) .. ($(a)+(-0.5,-0.9)$);
\draw[line width=(#3)*1.5pt,scale=#3] ($(a)+(0,1)$) .. controls ($(a)+(0.25,-0.5)$) .. ($(a)+(0.5,-0.9)$);
\draw[line width=(#3)*1.5pt,scale=#3] ($(a)+(0, -0.9)$) -- ($(a)+(-0.35, -0.7)$);
\draw[line width=(#3)*1.5pt,scale=#3] ($(a)+(0, -0.9)$) -- ($(a)+(0.35, -0.7)$);
\draw[line width=(#3)*0.75pt,scale=#3] ($(a)+(-0.35, -0.65)$) -- ($(a)+(0, -0.5)$);
\draw[line width=(#3)*0.75pt,scale=#3] ($(a)+(0.35, -0.65)$) -- ($(a)+(0, -0.5)$);
\draw[line width=(#3)*1pt,scale=#3] ($(a)+(0, -0.6)$) -- ($(a)+(-0.3, -0.45)$);
\draw[line width=(#3)*1pt,scale=#3] ($(a)+(0, -0.6)$) -- ($(a)+(0.3, -0.45)$);
\draw[line width=(#3)*0.5pt,scale=#3] ($(a)+(-0.3, -0.45)$) -- ($(a)+(0, -0.32)$);
\draw[line width=(#3)*0.5pt,scale=#3] ($(a)+(0.3, -0.45)$) -- ($(a)+(0, -0.32)$);
\draw[line width=(#3)*0.75pt,scale=#3] ($(a)+(0, -0.3)$) -- ($(a)+(-0.22, -0.17)$);
\draw[line width=(#3)*0.75pt,scale=#3] ($(a)+(0, -0.3)$) -- ($(a)+(0.22, -0.17)$);
\draw[line width=(#3)*0.5pt,scale=#3] ($(a)+(-0.22, -0.17)$) -- ($(a)+(0, -0.07)$);
\draw[line width=(#3)*0.5pt,scale=#3] ($(a)+(0.22, -0.17)$) -- ($(a)+(0, -0.07)$);;
\draw[line width=(#3)*0.75pt,scale=#3] (a) -- ($(a)+(-0.18, 0.11)$);
\draw[line width=(#3)*0.75pt,scale=#3] (a) -- ($(a)+(0.18, 0.11)$);
\draw[line width=(#3)*0.5pt,scale=#3] ($(a)+(-0.18, 0.11)$) -- ($(a)+(0,0.2)$);
\draw[line width=(#3)*0.5pt,scale=#3] ($(a)+(0.18, 0.11)$) -- ($(a)+(0,0.2)$);   
\draw[line width=(#3)*0.5pt,scale=#3] ($(a)+(0, 0.3)$) -- ($(a)+(-0.1, 0.37)$);
\draw[line width=(#3)*0.5pt,scale=#3] ($(a)+(0, 0.3)$) -- ($(a)+(0.1, 0.37)$);
\draw[line width=(#3)*0.25pt,scale=#3] ($(a)+(-0.1, 0.37)$) -- ($(a)+(0, 0.43)$);
\draw[line width=(#3)*0.25pt,scale=#3] ($(a)+(0.1, 0.37)$) -- ($(a)+(0, 0.43)$);
\draw[line width=(#3)*0.75pt,scale=#3] ($(a)+(0, 1.2)$) -- ($(a)+(0,1)$);
\draw[fill=white,scale=#3] ($(a)+(0, 1.2)$) circle (0.05cm);
\draw[line width=(#3)*1pt,decorate,decoration=expanding waves,scale=#3] ($(a)+(0, 1.2)$) -- ($(a)+(0, 2.5)$);
}
\newcommand{\basestationFullDuplex}[3]{
\coordinate (a) at (#1,#2);
\draw[line width=(#3)*1.5pt,scale=#3,color=red] ($(a)+(0, -1.08)$) -- ($(a)+(0, 1)$);
\draw[line width=(#3)*1.5pt,scale=#3,color=red] ($(a)+(0,1)$) .. controls ($(a)+(-0.25,-0.5)$) .. ($(a)+(-0.5,-0.9)$);
\draw[line width=(#3)*1.5pt,scale=#3,color=red] ($(a)+(0,1)$) .. controls ($(a)+(0.25,-0.5)$) .. ($(a)+(0.5,-0.9)$);
\draw[line width=(#3)*1.5pt,scale=#3,color=red] ($(a)+(0, -0.9)$) -- ($(a)+(-0.35, -0.7)$);
\draw[line width=(#3)*1.5pt,scale=#3,color=red] ($(a)+(0, -0.9)$) -- ($(a)+(0.35, -0.7)$);
\draw[line width=(#3)*0.75pt,scale=#3,color=red] ($(a)+(-0.35, -0.65)$) -- ($(a)+(0, -0.5)$);
\draw[line width=(#3)*0.75pt,scale=#3,color=red] ($(a)+(0.35, -0.65)$) -- ($(a)+(0, -0.5)$);
\draw[line width=(#3)*1pt,scale=#3,color=red] ($(a)+(0, -0.6)$) -- ($(a)+(-0.3, -0.45)$);
\draw[line width=(#3)*1pt,scale=#3,color=red] ($(a)+(0, -0.6)$) -- ($(a)+(0.3, -0.45)$);
\draw[line width=(#3)*0.5pt,scale=#3,color=red] ($(a)+(-0.3, -0.45)$) -- ($(a)+(0, -0.32)$);
\draw[line width=(#3)*0.5pt,scale=#3,color=red] ($(a)+(0.3, -0.45)$) -- ($(a)+(0, -0.32)$);
\draw[line width=(#3)*0.75pt,scale=#3,color=red] ($(a)+(0, -0.3)$) -- ($(a)+(-0.22, -0.17)$);
\draw[line width=(#3)*0.75pt,scale=#3,color=red] ($(a)+(0, -0.3)$) -- ($(a)+(0.22, -0.17)$);
\draw[line width=(#3)*0.5pt,scale=#3,color=red] ($(a)+(-0.22, -0.17)$) -- ($(a)+(0, -0.07)$);
\draw[line width=(#3)*0.5pt,scale=#3,color=red] ($(a)+(0.22, -0.17)$) -- ($(a)+(0, -0.07)$);;
\draw[line width=(#3)*0.75pt,scale=#3,color=red] (a) -- ($(a)+(-0.18, 0.11)$);
\draw[line width=(#3)*0.75pt,scale=#3,color=red] (a) -- ($(a)+(0.18, 0.11)$);
\draw[line width=(#3)*0.5pt,scale=#3,color=red] ($(a)+(-0.18, 0.11)$) -- ($(a)+(0,0.2)$);
\draw[line width=(#3)*0.5pt,scale=#3,color=red] ($(a)+(0.18, 0.11)$) -- ($(a)+(0,0.2)$);   
\draw[line width=(#3)*0.5pt,scale=#3,color=red] ($(a)+(0, 0.3)$) -- ($(a)+(-0.1, 0.37)$);
\draw[line width=(#3)*0.5pt,scale=#3,color=red] ($(a)+(0, 0.3)$) -- ($(a)+(0.1, 0.37)$);
\draw[line width=(#3)*0.25pt,scale=#3,color=red] ($(a)+(-0.1, 0.37)$) -- ($(a)+(0, 0.43)$);
\draw[line width=(#3)*0.25pt,scale=#3,color=red] ($(a)+(0.1, 0.37)$) -- ($(a)+(0, 0.43)$);
\draw[line width=(#3)*0.75pt,scale=#3,color=red] ($(a)+(0, 1.2)$) -- ($(a)+(0,1)$);
\draw[fill=white,scale=#3,color=red] ($(a)+(0, 1.2)$) circle (0.05cm);
\draw[line width=(#3)*1pt,decorate,decoration=expanding waves,scale=#3,color=red] ($(a)+(0, 1.2)$) -- ($(a)+(0, 2.5)$);
}

\newcommand{\RxTimeSharing}[3]{
\coordinate (a) at (#1,#2);
\draw[line width=0.25pt,scale=(#3)] (a)--($(a)+(-0.2,0)$)--($(a)+(-0.2,0.7)$)--
($(a)+(-0.1,0.8)$)--($(a)+(-0.3,0.8)$)--($(a)+(-0.2,0.7)$);
\draw ($(a)+(0.15,0)$) circle (0.15cm);
\node at ($(a)+(0.15,0)$) {\small +};
\draw ($(a)+(0.3,0)$)--($(a)+(0.5,0)$);
\draw[->] ($(a)+(0.15,-0.5)$)--($(a)+(0.15,-0.15)$);
\node at ($(a)+(0.15,-0.7)$) {$w$};
\draw[line width=0.25pt,scale=(#3)] ($(a)+(0.5,0)$)--($(a)+(0.6,0.6)+(0.5,0)$);
\draw[fill=black,scale=(#3)] ($(a)+(0.6,0)+(0.5,0)$) circle (0.03cm);
\draw[fill=black,scale=(#3)] ($(a)+(0.6,0.6)+(0.5,0)$) circle (0.03cm);
\draw[->]  ($(a)+(0.3,0.6)+(0.5,0)$) to[out=-20,in=70] ($(a)+(0.3,-0.1)+(0.5,0)$);
\draw[solid,scale=(#3)] ($(a)+(0.6,0)+(0.5,0)$)--($(a)+(1,0)+(0.5,0)$);
\draw[solid,scale=(#3)] ($(a)+(0.6,0.6)+(0.5,0)$)--($(a)+(1,0.6)+(0.5,0)$);
\draw[solid,scale=(#3)] ($(a)+(1,-0.2)+(0.5,0)$) rectangle ($(a)+(3,0.2)+(0.5,0)$);
\draw[solid,scale=(#3)] ($(a)+(1,0.4)+(0.5,0)$) rectangle ($(a)+(3,0.8)+(0.5,0)$);
\node[scale=(#3)] at ($(a)+(2,0.6)+(0.5,0)$){\small ID receiver};
\node[scale=(#3)] at ($(a)+(2,0)+(0.5,0)$){\small EH receiver};
}

\newcommand{\RxPowerSplitting}[3]{
\coordinate (a) at (#1,#2);
\draw[line width=0.25pt,scale=(#3)] (a)--($(a)+(-0.2,0)$)--($(a)+(-0.2,0.7)$)--
($(a)+(-0.1,0.8)$)--($(a)+(-0.3,0.8)$)--($(a)+(-0.2,0.7)$);
\draw ($(a)+(0.15,0)$) circle (0.15cm);
\node at ($(a)+(0.15,0)$) {\small +};
\draw ($(a)+(0.3,0)$)--($(a)+(0.5,0)$);
\draw[->] ($(a)+(0.15,-0.5)$)--($(a)+(0.15,-0.15)$);
\node at ($(a)+(0.15,-0.7)$) {$w$};
\draw ($(a)+(0,-0.2)+(0.5,0)$) rectangle ($(a)+(1,0.2)+(0.5,0)$);
\node at ($(a)+(0.5,0)+(0.5,0)$){\small PS};
\draw[line width=0.25pt,scale=(#3)]
($(a)+(1,-0.1)+(0.5,0)$)--($(a)+(1,-0.1)+(0.3,-0.3)+(0.5,0)$)
--($(a)+(1,-0.1)+(0.3,-0.3)+(0.8,0)+(0.5,0)$);
\draw[line width=0.25pt,scale=(#3)]
($(a)+(1,0.1)+(0.5,0)$)--($(a)+(1,0.1)+(0.3,0.3)+(0.5,0)$)
--($(a)+(1,0.1)+(0.3,0.3)+(0.8,0)+(0.5,0)$);
\draw[line width=0.25pt,scale=(#3)]
($(a)+(1,-0.1)+(0.3,-0.3)+(0.8,-0.2)+(0.5,0)$) rectangle 
($(a)+(1,-0.1)+(0.3,-0.3)+(2.8,0.2)+(0.5,0)$);
\draw[line width=0.25pt,scale=(#3)]
($(a)+(1,0.1)+(0.3,0.3)+(0.8,-0.2)+(0.5,0)$) rectangle
($(a)+(1,0.1)+(0.3,0.3)+(2.8,0.2)+(0.5,0)$);
\node at ($(a)+(1,0.1)+(0.3,0.3)+(1.8,0)+(0.5,0)$){\small ID receiver};
\node[scale=(#3)] at ($(a)+(1,-0.1)+(0.3,-0.3)+(1.8,0)+(0.5,0)$){\small EH receiver};
\node[scale=(#3)] at ($(a)+(1,0.1)+(0.3,0.3)+(0.3,0.15)+(0.5,0)$){\small $\sqrt{\eta}$};
\node[scale=(#3)] at ($(a)+(1,-0.1)+(0.3,-0.3)+(0.1,-0.2)+(0.5,0)$){\small $\sqrt{1-\eta}$};
}

\newcommand{\RxAntennaSeparation}[3]{
\coordinate (a) at (#1,#2);
\draw[line width=0.25pt,scale=(#3)] (a)--($(a)+(-0.2,0)$)--($(a)+(-0.2,0.7)$)--
($(a)+(-0.1,0.8)$)--($(a)+(-0.3,0.8)$)--($(a)+(-0.2,0.7)$);
\draw ($(a)+(0.15,0)$) circle (0.15cm);
\node at ($(a)+(0.15,0)$) {\small +};
\draw ($(a)+(0.3,0)$)--($(a)+(0.5,0)$);
\draw[->] ($(a)+(0.15,0.5)$)--($(a)+(0.15,0.15)$);
\node at ($(a)+(0.15,0.7)$) {$w$};
\draw[line width=0.25pt,scale=(#3)] ($(a)+(-0.2,-0.6)$)--($(a)+(-0.2,0)+(-0.2,-0.6)$)--
($(a)+(-0.2,0.7)+(-0.2,-0.6)$)--($(a)+(-0.1,0.8)+(-0.2,-0.6)$)--
($(a)+(-0.3,0.8)+(-0.2,-0.6)$)--($(a)+(-0.2,0.7)+(-0.2,-0.6)$);
\draw ($(a)+(0.15,0)+(-0.2,-0.6)$) circle (0.15cm);
\node at ($(a)+(0.15,0)+(-0.2,-0.6)$) {\small +};
\draw ($(a)+(0.3,0)+(-0.2,-0.6)$)--($(a)+(0.5,0)+(-0.2,-0.6)$);
\draw[->] ($(a)+(0.15,0)+(-0.2,-0.6)+(0,-0.5)$)
--($(a)+(0.15,0)+(-0.2,-0.6)+(0,-0.15)$);
\node at ($(a)+(0.15,0)+(-0.2,-0.6)+(0,-0.7)$) {$w$};
\draw ($(a)+(0.5,0)+(0,-0.2)$) rectangle
($(a)+(0.5,0)+(2,0.2)$);
\node at ($(a)+(0.5,0)+(1,0)$) {\small ID receiver};
\draw ($(a)+(0.5,0)+(-0.2,-0.6)+(0,-0.2)$) rectangle
($(a)+(0.5,0)+(-0.2,-0.6)+(2,0.2)$);
\node at ($(a)+(0.5,0)+(-0.2,-0.6)+(1,0)$) {\small EH receiver};
}
\newcommand{\Ali}{
\textcolor{black}
}
\begin{document}

\IEEEoverridecommandlockouts
\title{Heterogeneous Multi-Tier Networks:  Improper Signaling For Joint Rate-Energy Optimization}
	\author{{\small
		\IEEEauthorblockN{Ali Kariminezhad, \textit{Student Member, IEEE}, and Aydin Sezgin, \textit{Senior Member, IEEE}}}\\
		\thanks{{\small The authors are with the institute of Digital Communication Systems (DCS), Ruhr University Bochum (RUB), Bochum 44801, Germany. email: \{ali.kariminezhad, aydin.sezgin\}@rub.de}}
	}
\maketitle

\vspace{-1cm}
\begin{abstract}

Wireless nodes in future communication systems need to overcome three barriers when compared to their transitional counterparts, namely to support significantly higher data rates, have long-lasting energy supplies and remain fully operational in interference-limited heterogeneous networks. This could be partially achieved by providing three promising features, which are radio frequency (RF) energy harvesting, improper Gaussian signaling and operating in full-duplex communication mode, i.e., transmit and receive at the same time within the same frequency band.
In this paper, we consider these aspects jointly in a multi-antenna heterogeneous two-tier-network. In this network, the users in the femto-cell are sharing the scarce resources with the cellular users in the macro-cell and have to cope with the interference from the macro-cell base station as well as the transmitter noise and residual self-interference (RSI) due to imperfect full-duplex operation. Interestingly enough, while these impairments are detrimental from the achievable rate perspective, they are beneficial from the energy harvesting aspect as they carry RF energy. In this paper, we consider this natural trade-off jointly and propose appropriate optimization problems for beamforming and optimal resource allocation. Moreover, various receiver structures are employed for  both information detection (ID) and energy harvesting (EH) capabilities.
The paper aims at characterizing the trade-off between the achievable rates and harvested energies. Rate and energy maximization problems are thoroughly investigated. Finally, the numerical illustrations demonstrate the impact of energy harvesting on the achievable rate performance.
\end{abstract}

\begin{IEEEkeywords}
Heterogeneous networks, full-duplex communication, self-interference, energy harvesting, improper Gaussian signaling, Pareto boundary, augmented covariance matrix.
\end{IEEEkeywords}

\section{Introduction}
Wireless communication systems are facing difficulties in fulfilling the ever increasing demands of the customers operating in various communication standards. 
In order to fulfill the quality of service (QoS) demands of the users, the achievable rate region of the users need to be improved. Enhancing the achievable rates of the users with limited transmission power requires smart transceiver algorithms and techniques. Simultaneous transmission and reception within the same frequency band and time slot, i.e., full-duplex communications is an outstanding alternative for future communications, as it enables to almost doubling the spectral efficiency when compared to half-duplex counterpart. However, this comes not for free and additional hardware and processing is required to cancel the resulting self-interference due to the full-duplex operation~\cite{Bliss2014}. Self-interference can be partially suppressed passively by means of transmitter and receiver isolation \cite{Sabharwal2014, Shankar2012}, or it might be actively canceled in analog and digital domain by signal processing methods, \cite{Shankar20122, Bliss2012, Eltawil2015}. Thus, residual self-interference (RSI), which is assumed to be fully canceled in most theoretical works, still remains in practice. Moreover, transmitter noise due to non-linear behavior of the power amplifiers and limited dynamic range of the elements \cite{Hedley2008, Wichman2013} can not be ignored as well for such applications.
\begin{figure}
\centering
\tikzset{every picture/.style={scale=0.6}, every node/.style={scale=0.6}}%
\begin{tikzpicture}
    
    \draw[fill=yellow!20] (0,-0.5) ellipse (7cm and 2.75cm);

    \draw[fill=green!20] (3.7,0.5) ellipse (2cm and 1.1cm);
   \draw[fill=green!20] (-2.7,-0.5) ellipse (1.2cm and 1.2cm);
    \basestationFullDuplex{3.7}{0.5}{0.5}
    \draw[color=red,line width=1pt] (3.6,1.11) -- (3.8,1.11);
    \draw[color=red,line width=0.75pt] (3.6,1.11) -- (3.6,1.2);
    \draw[color=red,line width=0.75pt] (3.8,1.11) -- (3.8,1.2);
    \node[color=red] at (3.7,1.2) {..};
    \basestationHalfDuplex{0}{-0.15}{1.25}
    \draw[line width=1pt] (-0.50,1.1) -- (0.515,1.1);
    \draw[line width=0.75pt] (0.5,1.3) -- (0.5,1.1);
    \draw[line width=0.75pt] (-0.5,1.3) -- (-0.5,1.1);
    \draw[fill=white] (-0.5,1.34) circle (0.0625cm);
    \draw[fill=white] (0.5,1.34) circle (0.0625cm);
    \node at (0.25,1.3) {...};
    \node at (-0.25,1.3) {...};

    \iPhoneFullDuplex{4.75}{0.5}{0.5}

    \iPhoneFullDuplex{-2.8}{-1.2}{0.5}
    \iPhoneFullDuplex{-3.5}{-0.5}{0.5}
    \iPhoneFullDuplex{-2.2}{-1.4}{0.5}
    \iPhoneFullDuplex{-3}{-0.25}{0.5}
    \iPhoneHalfDuplex{-3.5}{-1.3}{0.5}


    \iPhoneHalfDuplex{2.5}{0}{0.5}
     
    \iPhoneHalfDuplex{-3.75}{0.75}{0.5}
    \iPhoneHalfDuplex{-1.75}{1.25}{0.5}
    \iPhoneHalfDuplex{-5.8}{-1}{0.5}

    \iPhoneHalfDuplex{2.6}{-2}{0.5}
    \iPhoneHalfDuplex{1.5}{1.2}{0.5}
    \iPhoneHalfDuplex{5.2}{-1}{0.5}
    \iPhoneHalfDuplex{1.7}{-0.9}{0.5}
    \basestationHalfDuplex{-8cm}{-4cm}{0.75};
    \draw[line width=0.6pt] (-8.4cm,-3.25cm) -- (-7.6cm,-3.25cm);
    \draw[line width=0.4pt] (-8.4cm,-3.25cm) -- (-8.4cm,-3.125cm);
    \draw[fill=white] (-8.4cm,-3.1cm) circle (0.035cm);
    \draw[line width=0.4pt] (-7.6cm,-3.25cm) -- (-7.6cm,-3.125cm);
    \draw[fill=white] (-7.6cm,-3.1cm) circle (0.035cm);
    \node at (-8.2cm,-3.15cm) {...};
    \node at (-7.8,-3.15cm) {...};
    \node at (-6.5cm,-4cm) {: Base Station};
    \basestationFullDuplex{-4.5cm}{-4cm}{0.5};
    \draw[color=red,line width=1pt] (-4.6,-3.4) -- (-4.4,-3.4);
    \draw[color=red,line width=0.75pt] (-4.6,-3.4) -- (-4.6,-3.3);
    \draw[color=red,line width=0.75pt] (-4.4,-3.4) -- (-4.4,-3.3);
    \node[color=red] at (-4.5,-3.3) {..};
    \node at (-2.9cm,-4cm) {: Access Point};
    \iPhoneHalfDuplex{-1.1cm}{-4.5cm}{0.75}
    \node at (0.75cm,-4cm) {: Mobile Users};
    \draw[fill=white] (3.5cm,-4cm) ellipse (1cm and 0.5cm);
    \node at (6cm,-4cm) {: Coverage areas};
\end{tikzpicture}
\caption{Full-duplex point-to-point (P2P) communication is performed in a femto-cell which is incorporated in a macro-cell. P2P communication can be performed by two mobile users in proximity, i.e., D2D communication.}
\label{fig:SysModelAll}
\end{figure}
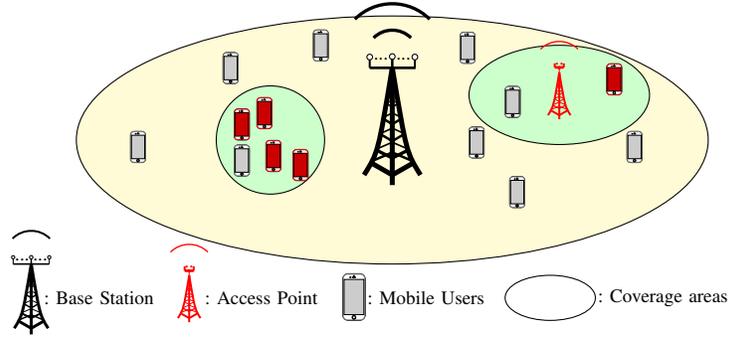
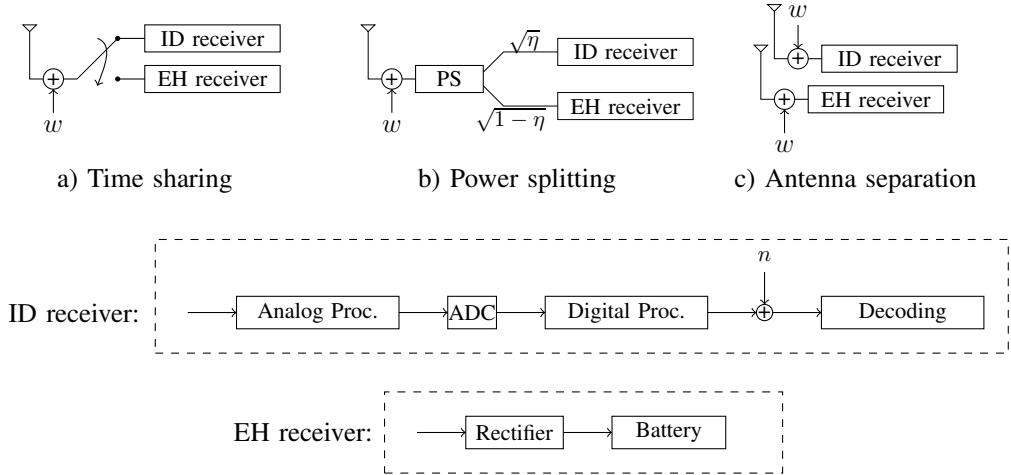
\begin{figure*}
\centering
\tikzset{every picture/.style={scale=.9}, every node/.style={scale=.85}}%
\begin{tikzpicture}
\RxTimeSharing{0}{0}{1}
\node at (1.5,-1.5) {a) Time sharing};
\RxPowerSplitting{5}{0}{1}
\node at (7,-1.5) {b) Power splitting};
\RxAntennaSeparation{11}{0.3}{1}
\node at (12,-1.5) {c) Antenna separation};
\end{tikzpicture}

\centering
\vspace*{0.5cm}
\tikzset{every picture/.style={scale=.9}, every node/.style={scale=.85}}%
\begin{tikzpicture}[scale=1.2]
\node at (-2,0){ID receiver:};
\draw [dashed] (-1,-0.5) rectangle (9.5,0.9);
\draw[->] (-0.6,0)--(0,0);
\draw (0,-0.2) rectangle (2,0.2);
\node at (1,0){\small Analog Proc.};
\draw[->] (2,0)--(2.6,0);
\draw (2.6,-0.2) rectangle (3.2,0.2);
\node at (2.9,0){\small ADC};
\draw[->] (3.2,0) -- (3.8,0);
\draw (3.8,-0.2) rectangle (5.8,0.2);
\node at (4.8,0){\small Digital Proc.};
\draw[->] (5.8,0) -- (6.4,0);
\draw (6.5,0) circle (0.1cm);
\node at (6.5,0){\small +};
\draw[->] (6.5,0.5)--(6.5,0.1);
\node at (6.5,0.7){\small $n$};
\draw[->] (6.6,0)--(7.2,0);
\draw (7.2,-0.2) rectangle (9.2,0.2);
\node at (8.2,0){\small Decoding};
\end{tikzpicture}

\vspace*{0.5cm}
\tikzset{every picture/.style={scale=.9}, every node/.style={scale=.85}}%
\begin{tikzpicture}[scale=1.2]
\node at (-2,0){EH receiver:};
\draw [dashed] (-1,-0.5) rectangle (3.9,0.5);
\draw[->] (-0.6,0)--(0,0);
\draw (0,-0.2) rectangle (1.2,0.2);
\node at (0.6,0){\small Rectifier};
\draw[->] (1.2,0) -- (1.8,0);
\draw (1.8,-0.2) rectangle (3.2,0.2);
\node at (2.5,0){\small Battery};
\end{tikzpicture}
\caption{Different schemes for EH and ID purposes. In order to study the performance of these schemes and be able to formulate optimization problems, we distribute the schemes among the users.}
\label{IdEh}
\vspace*{0.1cm}
\hrule
\end{figure*}
Accomplishing higher data rates with corresponding  signal processing tasks requires longer lasting energy supplies both for transmitters and receivers. Senders need to transmit with limited power due to hardware constraint (battery life-time) while the receivers are required to decode and process large amount of data under similar conditions. Hence, the users without the option for plug-in recharging, demand energy which needs to be provided in a wireless fashion. For this purpose, energy harvesting-capable (EHC) receivers could be deployed which harvest the energy from the environment, e.g. solar or RF energy. Thus, the life-time of the system can be improved from the energy in the air. The required energy is sometimes available at a user's surroundings and needs to be harvested, however, sometimes the required energy is not at its disposal and needs to be provided by the network. Thus, the study of power transmission and energy harvesting has become the focus of research community recently. For instance, the authors in~\cite{Wang2014} study delay-limited communication with EHC nodes. In~\cite{Ho2011}, the authors develop an outer bound for the rate-energy region considering energy harvesting constraints. Furthermore,~\cite{Gunduz2013} focuses on the sum rate optimization of an energy harvesting MISO communication system with feedback. The authors in~\cite{ZhangMay2013} study the performance limits of MIMO broadcast channel, in which the base station (BS) is responsible for both information and power transmission.\\
The above-mentioned results are valid for homogeneous networks. However,  practical communication systems are heterogeneous in nature, an aspect which has not been investigated so far. To this end, we consider a heterogeneous two-tier network with a single multiple-antenna macro-cell base station (BS) serving $K$ cellular users. Additionally, in a femto-cell, multiple pairs of multiple-antenna D2D nodes exchanges information in a full-duplex mode. Hence, the full-duplex D2D users suffer from both self-interference and interference from the cellular macro-cell users and vice versa. All users in this heterogeneous network, i.e., both cellular and full-duplex D2D users, are assumed to be equipped with an energy conversion chain that converts the incident RF signal energy to direct current in order to load the energy buffer, \cite{ZhouNov2013}. By this capability, the users' demand go beyond the traditional information transfer as they demand energy as well. Therefore, on one hand self-interference and the interference from the other users is deteriorating the process of decoding the desired signal reliably, on the other hand the users could use the energy of the interference for EH purposes. Considering energy and information rate demands of the users, we study the performance limits of the cellular and D2D users in the network. These limits are due to the intrinsic trade-off between the demands. Considering this trade-off, the optimal rate tuples of the cellular users and full-duplex D2D users capable of EH are studied. Moreover, the optimal rate-energy pairs are investigated. Thus, we address two main problems,
\begin{itemize}
\item What are the achievable rate region of the cellular users and D2D pair under certain transmit power and received energy constraints?
\item What are the optimal rate-energy tuples of the D2D users under cellular users' QoS and power constraints? 
\end{itemize}
The two questions will be answered in an optimization framework. We will establish appropriate optimization problems for joint information detection (ID) and EH transceiver structures and compare their performance. In this work, different ID and EH receivers are investigated. The users could be equipped with antenna separation (AS) receivers, where the energy and information of RF signals are caught simultaneously over different antennas. Power splitting (PS) and time-sharing (TS) are other alternatives for joint ID and EH purposes~\cite{BiApril2015}. By splitting the received signal power, the energy of one portion is converted to direct current for loading the energy buffer, while the information out of the other portion is decoded~\cite{Liu2013,Kariminezhad2017SPL}. Time-sharing between energy harvesting and information detection phases allows EH and ID in separate time instants, Fig.\ref{IdEh}.\\ 
Furthermore, within this context we compare proper Gaussian signaling with improper Gaussian signaling~\cite{Jafar2009} in the transmission phase. Improper Gaussian signaling has been shown to be beneficial in interference channels (IC) and X-channels from the achievable rate and consumed power perspectives~\cite{Jafar2009,Jorswieck2012,Zhang2013,Guan2013,Kariminezhad2017A}.\\
By utilizing improper Gaussian signaling, the outermost rate region and rate-energy region is investigated by formulating Chebyshev weighting function, \cite{Ottersten2014}. Then, the problems are reformulated as semi-definite programs (SDP) with non-convex rank-1 constraints. These constraints are relaxed and the resulting semidefinite relaxation (SDR) is solved efficiently, \cite{Luo2010}. If the optimal solutions are not rank-1, the Gaussian randomization process ~\cite{Quan2006,Sidi2006} is further utilized to acquire sub-optimal rank-1 solutions.\\
\textit{Contribution}: To sum up, in this paper we utilize rather novel improper Gaussian signaling in a two-tier network with full-duplex communication. Hence, we generalize the transmitter noise model of the full-duplex users for this type of Gaussian transmission. Furthermore, the (self)-interference in the network is proposed to be harvested in order to load the energy buffer instead of being wasted. The rate-region and rate-energy region are studied and the performance of improper Gaussian signaling is compared with the proper Gaussian signaling. Moreover, non-linear precoding is compared with the utilized widely linear solutions.
\section{System Model}
In this paper, we  consider a cellular network as shown in Fig.~\ref{fig:SysModelAll} in which a base-station equipped with $N$ antennas is serving a set of $K$ cellular users. This network operates in a half-duplex mode, i.e., the uplink and downlink operation is performed in successive time instances. In order to overcome the limitations of their local battery supplies, the cellular users are equipped with energy-harvesting (EH) receiver chains. Those receiver chains capture the energy of the RF signals in their environment. Furthermore, in this cell several pairs of D2D users are deployed, which exchange data in a full-duplex mode, i.e., the D2D users are able to receive and transmit at the time within the same frequency band. Here, we follow the design proposed and utilized in~\cite{Gorokhov2004},~\cite{Gore2002}, in which a full-duplex node is using a subset $M$ of its antennas for transmission and the remaining ones for reception. Similar to the cellular users, the D2D users are equipped with EH receiver chains. 

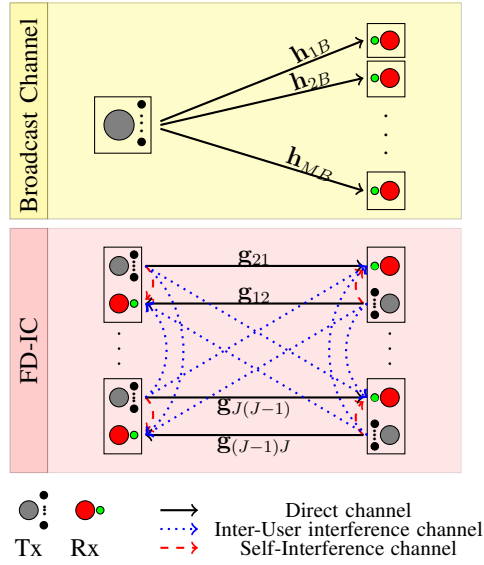
\begin{figure}[t]
\centering
\tikzset{every picture/.style={scale=.5}, every node/.style={scale=.8}}%
\begin{tikzpicture}
\draw[fill=yellow, opacity=0.3] (-3,1) rectangle (-2,-4.75);\node[rotate=90] at (-2.5,-2) {Broadcast Channel};
\draw[fill=yellow, opacity=0.2] (-2,1) rectangle (9,-4.75);

\draw[fill=red, opacity=0.2] (-3,-5) rectangle (-2,-8-3.5);\node[rotate=90] at (-2.5,-8.25) {FD-IC};
\draw[fill=red,opacity=0.1] (-2,-5) rectangle (9,-8-3.5);
\draw[fill=gray] (-0.1,-2.25) circle (0.4cm);
\draw (-0.75,-3) rectangle (0.75,-1.5);
\draw[fill=black] (0.5,-1.7) circle (0.1cm);
\node at (0.5,-2) {.};
\node at (0.5,-2.2) {.};
\node at (0.5,-2.4) {.};
\draw[fill=black] (0.5,-2.7) circle (0.1cm);
\draw (-.5+7,-.45) rectangle (.5+7,.45);
\draw[fill=red] (7.1,0) circle (0.25cm);
\draw[fill=green] (6.7,0) circle (0.1cm);

\draw[->,thick] (1,-2.15) to (6.4,0);
\node[rotate=30] at (5,-0.2) {${\bf h}_{1B}$};
\draw (-.5+7,-.5-1) rectangle (.5+7,.5-1.05);
\draw[fill=red] (7.1,-1) circle (0.25cm);
\draw[fill=green] (6.7,-1) circle (0.1cm);

\draw[->,thick] (1,-2.25) to (6.4,-1);
\node[rotate=20] at (5,-1) {${\bf h}_{2B}$};

\draw (-.5+7,-.5-4) rectangle (.5+7,.5-4);
\draw[fill=red] (7.1,-4) circle (0.25cm);

\draw[fill=green] (6.7,-4) circle (0.1cm);

\draw[->,thick] (1,-2.35) to (6.4,-4);
\node[rotate=-20] at (5,-3.3) {${\bf h}_{MB}$};
\node at (7,-2) {.};
\node at (7,-2.5) {.};
\node at (7,-3) {.};

\draw (-.5,-7.5) rectangle (0.5,-5.5);

\draw[fill=gray] (-0.1,-6) circle (0.25cm);
\draw[fill=black] (0.3,0.3-6) circle (0.1cm);
\node at (0.3,0.1-6) {.};
\node at (0.3,0.0-6) {.};
\node at (0.3,-0.1-6) {.};
\draw[fill=black] (0.3,-0.3-6) circle (0.1cm);

\draw[fill=red] (-0.1,-7) circle (0.25cm);
\draw[fill=green] (0.3,0.3-6-1.3) circle (0.1cm);
\draw[->,color=red,dashed,thick] (0.6,-6) to [bend left=50] (0.6,-6.9);
\draw (-.5+7,-7.5) rectangle (0.5+7,-5.5);

\draw[fill=red] (7+0.1,-6) circle (0.25cm);
\draw[fill=green] (7+0.1-0.4,-6) circle (0.1cm);\fill[color=green] (7+0.1-0.4-0.1,-6) arc (180:360:0.1cm);

\draw[fill=gray] (7+0.1,-7) circle (0.25cm);
\draw[fill=black] (7+0.3-0.6,0.3-6-1) circle (0.1cm);
\node at (7+0.3-0.6,0.1-6-1) {.};
\node at (7+0.3-0.6,0.0-6-1) {.};
\node at (7+0.3-0.6,-0.1-6-1) {.};
\draw[fill=black] (7+0.3-0.6,-0.3-6-1) circle (0.1cm);
\draw[->,color=red,dashed,thick] (6.4,-7) to [bend left=50] (6.4,-6.1);
\draw[->,thick] (0.6,-6) to (6.4,-6);
\node at (3.5,-5.8) {${\bf g}_{21}$};
\draw[->,thick] (6.4,-7) to (0.6,-7);
\node at (3.5,-6.8) {${\bf g}_{12}$};
%

\draw (-0.1,-7.8) circle (0.02cm);
\draw (-0.1,-8.2) circle (0.02cm);
\draw (-0.1,-8.6) circle (0.02cm);

\draw (7.1,-7.8) circle (0.02cm);
\draw (7.1,-8.2) circle (0.02cm);
\draw (7.1,-8.6) circle (0.02cm);

\draw (-.5,-7.5-3.5) rectangle (0.5,-5.5-3.5);

\draw[fill=gray] (-0.1,-6-3.5) circle (0.25cm);
\draw[fill=black] (0.3,0.3-6-3.5) circle (0.1cm);
\node at (0.3,0.1-6-3.5) {.};
\node at (0.3,0.0-6-3.5) {.};
\node at (0.3,-0.1-6-3.5) {.};
\draw[fill=black] (0.3,-0.3-6-3.5) circle (0.1cm);

\draw[fill=red] (-0.1,-7-3.5) circle (0.25cm);
\draw[fill=green] (0.3,0.3-6-1.3-3.5) circle (0.1cm);
\draw[->,color=red,dashed,thick] (0.6,-6-3.5) to [bend left=50] (0.6,-6.9-3.5);

\draw (-.5+7,-7.5-3.5) rectangle (0.5+7,-5.5-3.5);

\draw[fill=red] (7+0.1,-6-3.5) circle (0.25cm);
\draw[fill=green] (7+0.1-0.4,-6-3.5) circle (0.1cm);\fill[color=green] (7+0.1-0.4-0.1,-6-3.5) arc (180:360:0.1cm);

\draw[fill=gray] (7+0.1,-7-3.5) circle (0.25cm);
\draw[fill=black] (7+0.3-0.6,0.3-6-1-3.5) circle (0.1cm);
\node at (7+0.3-0.6,0.1-6-1-3.5) {.};
\node at (7+0.3-0.6,0.0-6-1-3.5) {.};
\node at (7+0.3-0.6,-0.1-6-1-3.5) {.};
\draw[fill=black] (7+0.3-0.6,-0.3-6-1-3.5) circle (0.1cm);
\draw[->,color=red,dashed,thick] (6.4,-7-3.5) to [bend left=50] (6.4,-6.1-3.5);
\draw[->,thick] (0.6,-6-3.5) to (6.4,-6-3.5);
\node at (3.5,-5.8-3.5-0.5) {${\bf g}_{J(J-1)}$};
\draw[->,thick] (6.4,-7-3.5) to (0.6,-7-3.5);
\node at (3.5,-6.8-3.5-0.5) {${\bf g}_{(J-1)J}$};
%
\draw[->,dotted,blue,thick] (0.6,-6) to [bend left=50] (0.6,-6-1-3.5);
\draw[->,dotted,blue,thick] (0.6,-6) to  (6.5,-6-3.5);

\draw[->,dotted,blue,thick] (0.6,-6-3.5) to [bend left=-50] (0.6,-6-1);
\draw[->,dotted,blue,thick] (0.6,-6-3.5)  to (6.5,-6);

\draw[->,dotted,blue,thick] (6.5,-6-1) to [bend left=-50] (6.5,-6-3.5);
\draw[->,dotted,blue,thick] (6.5,-6-1) to  (0.6,-6-1-3.5);

\draw[->,dotted,blue,thick] (6.5,-6-3.5-1) to [bend left=50] (6.5,-6);
\draw[->,dotted,blue,thick] (6.5,-6-3.5-1) to  (0.6,-6-1);


\draw[fill=gray] (-2.5,-9-3.5) circle (0.25cm);
\draw[fill=black] (-2.5+0.4,-9+0.4-3.5) circle (0.1cm);
\node at (-2.5+0.425,-9+0.1-3.5){.};
\node at (-2.5+0.425,-9+0-3.5){.};
\node at (-2.5+0.425,-9-0.1-3.5){.};
\draw[fill=black] (-2.5+0.4,-9-0.4-3.5) circle (0.1cm);
\node at (-2.5,-10-3.5) {Tx};

\draw[fill=red] (-1,-9-3.5) circle (0.25cm);
\draw[fill=green] (-1+0.4,-9-3.5) circle (0.1cm);
\node at (-1,-10-3.5) {Rx};

\draw[color=black,->,thick] (1,-9-3.5)--(2,-9-3.5);
\node at (6,-9-3.5) {\small Direct channel};
\draw[color=blue,->,dotted,thick] (1,-9.5-3.5)--(2,-9.5-3.5);
\node at (6,-9.5-3.5) {\small Inter-User interference channel};
\draw[color=red,->,dashed,thick] (1,-10-3.5)--(2,-10-3.5);
\node at (6,-10-3.5) {\small Self-Interference channel};

\end{tikzpicture}
\caption{A two-tier network with cellular and D2D full-duplex communications. The base station and D2D users are equipped with multiple antennas. Note that the interference between two tiers is not depicted for clarity in illustration.}
\label{fig:SysModel}
\end{figure}

Now, let the set of cellular users be denoted as $\mathcal{C}$. For convenience, we define the set of D2D users as $\mathcal{D}$. Furthermore, the number of cellular and D2D users are defined as $K=|\mathcal{C}|$ and $J=|\mathcal{D}|$, respectively. Then,  the channel input-output relationships at each time instant (we skip the time index) are given by
{\small
\begin{align}
y_k=&
\mathbf{h}_{kB}^H({\bf x}_B+{\bf e}_B)+\sum_{j=1}^{J}{\bf h}_{kj}^H ({\bf x}_j+{\bf e}_j)+w_k +n_k, \quad \forall k\in\mathcal{C},\quad \text{(cellular)},\\
z_j=&
{\bf g}_{jB}^H ({\bf x}_B+{\bf e}_B)+\sum_{\substack{i=1\\i\neq j}}^{J}{\bf g}_{ji}^H ({\bf x}_i+{\bf e}_i)+ {\bf g}_{jj}^H ({\bf x}_j + {\bf e}_j) +w^{'}_j+ n^{'}_j, \quad \forall j\in\mathcal{D},\quad \text{(full-duplex)},
\end{align}}
where $y_{k}$ and $z_{j}$ denote the received signals at the $k$th cellular user and at $j$th D2D user, respectively. Furthermore, the transmitter noise is expressed by ${\bf e}\in\mathbb{C}^{ M \times 1}$. Transmitter noise appears mainly due to the limited transmitter dynamic range (DR).
The entities $n$ and $n^{'}$ represent  realizations of independent and identically distributed zero-mean proper Gaussian noise with variance $\sigma^{2}_n$, i.e., $\mathcal{CN}(0,\sigma^{2}_n)$. This noise is due to the imperfections in the ID receiver chain (e.g., phase noise, thermal noise and quantization noise). Antenna noise $w_k$ and $w^{'}_j$ are modeled as zero-mean proper AWGN with variance $\sigma^{2}_w$, i.e., $\mathcal{CN}(0,\sigma^{2}_w)$. The interference channel vector between the $j$th D2D user and the BS is denoted by $\mathbf{g}_{jB}\in\mathbb{C}^{ N \times 1}$ and the self-interference channels are represented by $\mathbf{g}_{jj}\in\mathbb{C}^{ M \times 1}$. The direct link between the $j$th and $i$th D2D users is given by $\mathbf{g}_{ji}\in\mathbb{C}^{ M \times 1}$. The channel vectors from the BS and the $j$th D2D user to the $k$th cellular user are represented by $\mathbf{h}_{kB}\in\mathbb{C}^{ N \times 1}$ and $\mathbf{h}_{kj}\in\mathbb{C}^{ M \times 1}$, respectively. The system model with the respective channels between the users is shown in~Fig. (\ref{fig:SysModel}).

The transmit signal of the BS is denoted as $\mathbf{x}_B\in\mathbb{C}^{N\times 1}$ which is given by
{\small
\begin{align}
{\bf x}_B=&\sum\limits_{k=1}^{N} {\bf x}_{B_k} =\sum\limits_{k=1}^{N} {\bf v}_{B_k} d_{B_k}={\bf V}_B{\bf d}_{B},\label{BF1}
\end{align}}
where $d_{B_k}$ and ${\bf v}_{B_k}$ are the $k$th information symbol and beamforming vector intended for the $k$th cellular user, respectively. The BS transmit beamforming matrix ${\bf V}_B$ and the transmit information signal vector ${\bf d}_{B}$ are defined as ${\bf V}_B=[{\bf v}_{B_1},...,{\bf v}_{B_K} ]$ and ${\bf d}_{B}=[d_{B_1},...,d_{B_K}]^{T}$, respectively.
Similarly, the transmit signal of the D2D users is given by
{\small
\begin{align}
{\bf x}_{j}=&{\bf v}_{j} d_{j},\quad \forall j\in\mathcal{D},\label{BF2}
\end{align}}
where the information signal $d_j$ is beamformed in the direction of ${\bf v}_{j}$. Note that the information symbols $d_{B_k},\ d_{j},\  \forall k\in\mathcal{C},\ j\in\mathcal{D}$ are assumed to be independently identically distributed complex Gaussian with unit variance. If the real and imaginary parts of $d_{B_k},\ \forall k\in\mathcal{C}$ and $d_{j},\ \forall j\in\mathcal{D}$ have equal power and are uncorrelated, then the signaling type is referred to as proper Gaussian signaling. Otherwise, it is referred to as improper Gaussian signaling~\cite{Schreier2010}.\\
Moreover, we assume here that the D2D users are equipped with $M+1$ antennas, where $M$ antennas are utilized for transmission and a single antenna is used for reception. 

In this work, we assume perfect and global channel knowledge. The self-interference due to full-duplex operation is assumed to be canceled to some significant extent (based on the SI channel knowledge), but not completely (due to the transmitter noise). Thus, we rewrite the received signals at the D2D users as
{\small
\begin{align}
\hat{z}_j=
{\bf g}_{jB}^H ({\bf x}_B+{\bf e}_B)+\sum_{\substack{i=1\\i\neq j}}^{J}{\bf g}_{ji}^H ({\bf x}_i+{\bf e}_i)
+ {\bf g}_{jj}^H{\bf e}_j + w^{'}_j + n^{'}_j, \quad \forall j\in\mathcal{D},\nonumber
\end{align}}
where the residual self-interference (RSI) due to transmitter noise is represented by ${\bf g}_{jj}^H{\bf e}_j$. Assuming improper Gaussian signaling, the transmitter noise $\mathbf{e}_j$ is modeled as
{\small
\begin{align}
\quad{\bf e}_j&\indep{\bf x}_j,\quad \Ali{{\bf e}_j\sim\mathcal{N}({\bf 0} , \tilde {\bf Q}_{e_j})},\label{TXnoiseInd}\\
\tilde {\bf Q}_{e_j}&=\kappa\tilde {\bf C}_{x_j}=\kappa\begin{bmatrix}
\mathbf{C}_{x_j} & \hat{\mathbf{C}}_{x_j}\\
\hat{\mathbf{C}}^{*}_{x_j} & 
\mathbf{C}^{*}_{x_j}
\end{bmatrix},\label{TXnoiseInd2}
\end{align}}
which states that the transmitter noise follows an improper Gaussian distribution with zero mean and augmented covariance matrix $\kappa\tilde {\bf C}_{x_j}$ with $\kappa\ll 1$. Notice that, the transmit signal augmented covariance matrix $\tilde {\bf C}_{x_j}$ consists of the signal covariance, i.e.,  $\mathbf{C}_{x_j}=\mathbb{E}\{\mathbf{x}_j\mathbf{x}^{H}_j\}$, and pseudo-covariance, i.e.,  $\hat{\mathbf{C}}_{x_j}=\mathbb{E}\{\mathbf{x}_j\mathbf{x}^{T}_j\}$, matrices. Hence, it characterizes the second-order moment thoroughly. As given by (\ref{TXnoiseInd}), the transmitter noise is statistically independent from the transmit signal. The assumption of an improper transmitter noise is due to the generated improper information signal in baseband and imbalance between the in-phase and quadrature (I/Q) components, where the latter is discussed in~\cite{Adali2011}. The authors in \cite{Bliss2012} propose a transmitter noise model whose covariance is composed of the diagonals of the transmit signal covariance matrix. By plugging their model in our general model, (\ref{TXnoiseInd2}) is recast as
{\small
\begin{align}
\tilde {\bf Q}_{e_j}&=\kappa\tilde {\bf C}_{x_j}=\kappa\begin{bmatrix}
{\rm diag}(\mathbf{C}_{x_j}) & \hat{\mathbf{C}}_{x_j}\\
\hat{\mathbf{C}}^{*}_{x_j} & 
{\rm diag}(\mathbf{C}^{*}_{x_j})
\end{bmatrix}.\label{TXnoiseInd3}
\end{align}}
The transmitter noise undergoes self-interference channel and can not be canceled at the receiver. This is due to the absence of transmitter noise knowledge at the receivers. However, except for the D2D users the contribution of the transmitter noise can be ignored at all receivers. This assumption is valid due to the low power of transmitter noise and relative strength of self-interference channel compared to other channels. Hence, the system model is simplified to
{\small
\begin{align}
\Ali{y_k=}&
\Ali{{\bf h}^{H}_{kB}{\bf x}_B+\sum_{j=1}^{J}{\bf h}_{kj}^{H} {\bf x}_j+ w_k+ n_k, \quad \forall k\in\mathcal{C},}\label{ss1}\\
\Ali{\hat{z}_j=}&
\Ali{{\bf g}_{jB}^H {\bf x}_B+\sum_{\substack{i=1\\i\neq j}}^{J}{\bf g}_{ji}^H {\bf x}_i
+{\bf g}_{jj}^{H} {\bf e}_j+ w^{'}_j + n^{'}_j, \quad \forall j\in\mathcal{D}.}\label{ss2}
\end{align}}
By plugging (\ref{BF1}) and (\ref{BF2}) into (\ref{ss1}) and (\ref{ss2}), the received signals are recast as
{\small
\begin{align}
y_k=&
\underbrace{{\bf h}_{kB}^{H}\sum_{\substack{m=1\\ m\neq k}}^M {\bf v}_{B_m} d_{B_m}+\sum_{j=1}^{J}{\bf h}_{kj}^{H}{\bf v}_j d_{j}}_{\text{interference}}+\underbrace{{\bf h}^{H}_{kB}{\bf v}_{B_k} d_{B_k}}_{\text{desired}}
+w_k+n_k,\quad \forall k\in\mathcal{C},\label{SysMod1}\\
\hat{z}_j=&
\underbrace{{\bf g}^{H}_{jB} \sum\limits_{m=1}^M {\bf v}_{B_m} d_{B_m} +\sum_{\substack{i=1\\i\neq j,l\\ }}^{J}{\bf g}_{ji}^H {\bf v}_i d_{i}+{\bf g}_{jj}^{H} {\bf e}_j}_\text{interference}+\underbrace{{\bf g}_{jl}^{H}{\bf v}_{l} d_{l}}_{\text{desired}}+w^{'}_j + n^{'}_j, \quad \forall j\in\mathcal{D}.\label{SysMod2}
\end{align}}

Here, we observe the dilemma we are facing in harvesting energy in this network. While the interference terms in expressions (\ref{SysMod1}) and (\ref{SysMod2}) are detrimental to the rate performance as they represent harmful interference, they are beneficial for energy harvesting as they posses energy. Note that index $l$ in \eqref{SysMod2} is defined as
{\small
\begin{align}
l=\begin{cases}
j-1,\quad j\in\mathcal{D}_e=\{\mathcal{D}\bigcap \mathbb{N}_e\}\\
j+1,\quad j\in\mathcal{D}_o=\{\mathcal{D}\bigcap \mathbb{N}_o\}
\end{cases},
\end{align}}
where $\mathbb{N}_e$ and $\mathbb{N}_o$ are the set of even and odd natural numbers, respectively.\\
In this work, we investigate various types of ID and EH chains at the receiver that will be discussed in the following.
We utilize the models introduced in \cite{BiApril2015} for simultaneous wireless information and energy reception. For the purpose of ID, both cellular and full-duplex users deploy single receive antenna. For the purpose of EH, different structures are utilized,
\begin{itemize}
\item Antenna separation (AS): The users could be equipped with an extra receive antenna for EH purpose. We assume that the signals arriving at both antennas (one for ID and one for EH) are experiencing fully-correlated channels. Due to the small-size hand-held mobile stations, the physical distance between the antenna elements in an array antenna is small. Thus the received signals are highly correlated.
\item Power splitting (PS): The users could split received signal power for joint ID and EH in one channel use. This could be achieved by utilizing a power splitter at the receivers.
\item Time sharing (TS): The users have the option to change the receive strategy and do time-sharing between ID and EH phases, (ID and EH in different channel uses).
\end{itemize}
For simplicity in presenting the optimization problems, we distribute the aforementioned joint ID and EH techniques among the users. We allow cellular users to harvest the energy of the incident RF signal by AS structure, while full-duplex D2D users employ either PS or TS for energy harvesting purpose.   
With this energy harvesting receivers for the users in the network, we will formulate the achievable rates and energies in the next section.
\section{Achievable Rates and Energies}
In this section, we formulate the achievable rates of the users assuming Gaussian codebook at the transmitters. In order to decode the desired signals, the users ignore interference, i.e., treat interference as noise (TIN). The cellular and full-duplex users' achievable rates are bounded by
{\small
\begin{align}
r_k&\leq I(y_k;{\bf x}_{B_k})=h({y}_k)-h(y_k|{\bf x}_{B_k}),\quad\forall k\in\mathcal{C},\label{Rr11} \\
r^{'}_j&\leq I(\hat{z}_j;{\bf x}_l)=h(\hat{z}_j)-h(\hat{z}_j|{\bf x}_{l}),\quad\forall j\in\mathcal{D} \label{Rr22},
\end{align}}
respectively, where $I(y_k;{\bf x}_{B_k})$ is the mutual information between $y_k$ and ${\bf x}_{B_k}$ and $h({y}_k)$ is the differential entropy of $y_k$,~\cite{Cover1991}. Moreover, the differential entropy of $y_k$ given ${\bf x}_{B_k}$ is $h(y_k|{\bf x}_{B_k})$.\\
{\it{Definition:}}
The differential entropy of a complex Gaussian random variable $y_k$ is given by \cite{Schreier2010},
{\small
\begin{align}
h(y_k)= \frac{1}{2}\log\left((2\pi e)^{2}|\tilde{\bf C}_{y_k}|\right),\label{entropyAsym}
\end{align}}
where $y_k\in\mathbb{C}$. For the case of proper Gaussian where $\hat{\bf C}_{y_k}=0$, the differential entropy expression reduces to $h({\bf x})= \log(2\pi eC_{y_k})$.\\
Now, by plugging~\eqref{entropyAsym} into~\eqref{Rr11} we obtain
{\small
\begin{align}
r_k&\leq \frac{1}{2}\log\left(\frac{|{\tilde{\bf C}}_{y_k}|}{| {\tilde{\bf C}}_{w_k}|}\right)=\frac{1}{2}\log\left(\frac{C^{2}_{y_k}-|\hat C_{y_k}|^{2}}{C^{2}_{w_k}-|\hat C_{w_k}|^{2}}\right)\nonumber\\
&=\underbrace{\log\left(\frac{C_{y_k}}{C_{w_k}}\right)}_{R_k^{\mathsf{proper}}}+
\underbrace{\frac{1}{2}\log\left(\frac{1-C^{-2}_{y_k}|\hat C_{y_k}|^{2}}{1-C^{-2}_{w_k}|\hat C_{w_k}|^{2}}\right)}_{R_k^{\mathsf{improper}}}:=R_k,\,\ \forall k\in\mathcal{C},\label{Rr1}
\end{align}}
where $C_{w_k}$  and $\hat C_{w_k}$ are the variance and pseudo-variance of the interference-plus-noise at the $k$th cellular user. Notice that, the first terms in (\ref{Rr1}) and (\ref{Rr2}) correspond to the achievable rate bound in case of proper signaling, i.e., $\hat C_{y_k}=0$ and $\hat C_{z_j}=0$. Similarly, the achievable rate of the $j$th full-duplex user is given by
{\small
\begin{align}
r^{'}_j&\leq\frac{1}{2}\log\left(\frac{|{\tilde{\bf C}}_{z_j}|}{|{\tilde{\bf C}}_{q_j}|}\right)=\frac{1}{2}\log\left(\frac{C^{2}_{z_j}-|\hat C_{z_j}|^{2}}{C^{2}_{q_j}-|\hat C_{q_j}|^{2}}\right),\nonumber\\
&=\underbrace{\log\left(\frac{C_{z_j}}{C_{q_j}}\right)}_{R_j^{'{\mathsf{proper}}}}+
\underbrace{\frac{1}{2}\log\left(\frac{1-C^{-2}_{z_j}|\hat C_{z_j}|^{2}}{1-C^{-2}_{q_j}|\hat C_{q_j}|^{2}}\right)}_{R_j^{'{\mathsf{improper}}}}:=R^{'}_j \quad \forall j\in\mathcal{D} \label{Rr2},
\end{align}}
where $C_{q_j}$  and $\hat C_{q_j}$ are the variance and pseudo-variance of the interference-plus-noise at the $j$th full-duplex user. Allowing the transmission to be improper Gaussian, we can enhance the bound by improving the second terms in (\ref{Rr1}) and (\ref{Rr2}), \cite{Jorswieck2012,Zhang2013}.\\
Here, we define the received signals and interference-plus-noise variances and pseudo-variances that are required in~\eqref{Rr1} and~\eqref{Rr2}. The variance of the received signals at the $k$th cellular and $j$th full-duplex users are formulated as 
{\small
\begin{align}
C_{y_k}=&
{\bf h}_{kB}^{H}{\bf C}_{x_{B}}{\bf h}_{kB}+\sum_{j=1}^{J}{\bf h}^{H}_{kj}{\bf C}_{x_j}{\bf h}_{kj}+\sigma _w^{2}+\sigma _n^{2},\quad \forall k\in\mathcal{C},\label{cy1}\\
C_{z_j}=&
{\bf g}_{jB}^{H}{\bf C}_{x_B}{\bf g}_{jB}+\sum_{\substack{i=1\\i\neq j }}^{J}{\bf g}_{ji}^H{\bf C}_{x_i}{\bf g}_{ji}+
\kappa{\bf g}_{jj}^{H}{\rm diag}({\bf C}_{x_j}){\bf g}_{jj}+\sigma_w^{2}+\sigma_n^{2}, \quad \forall j\in\mathcal{D}, \label{cz1}
\end{align}}
respectively, where ${\bf C}_{x_B}={\bf V}_B
\mathbb{E}\{{\bf d}_B{\bf d}^{H}_B\}{\bf V}^{H}_B$ and ${\bf C}_{x_j}={\bf v}_j
\mathbb{E}\{ d_j d^{H}_j\}{\bf v}^{H}_j$ are the BS and D2D transmit covariance matrices, respectively. 
Moreover, we formulate the interference-plus-noise variance as
{\small
\begin{align}
C_{w_k}&=
C_{y_k}-{\bf h}^{H}_{kB}{\bf C}_{x_{B_k}}{\bf h}_{kB},\quad \forall k\in\mathcal{C},\label{cw1}\\
C_{q_j}&=
C_{z_j}-{\bf g}^{H}_{jl}{\bf C}_{x_{l}}{\bf g}_{jl}, \quad \forall j\in\mathcal{D},
\label{cq1}
\end{align}}
where ${\bf C}_{x_{B_k}}={\bf v}_{B_k}
\mathbb{E}\{ d_{B_k} d^{*}_{B_k}\}{\bf v}^{H}_{B_k}$ is the $k$th cellular user's desired stream covariance matrix.
In addition to the variances, the pseudo-variances of the received signals and interference-plus-noise are required in order to obtain the augmented covariance matrices required in the rate expressions in (\ref{Rr1}) and (\ref{Rr2}) .We write the pseudo-variance of the received signal as
{\small
\begin{align}
\hat C_{y_k}=&
{\bf h}^{H}_{kB}{\bf\hat C}_{x_B}{\bf h}^{*}_{kB}+\sum_{j=1}^{J}{\bf h}^{H}_{kj}{\bf\hat C}_{x_j}{\bf h}^{*}_{kj},\quad \forall k\in\mathcal{C},\label{chaty1}\\
\hat C_{z_j}=&
{\bf g}_{jB}^{H}{\bf\hat C}_{x_B}{\bf g}^{*}_{jB}+\sum_{\substack{i=1\\i\neq j }}^{J}{\bf g}_{ji}^H{\bf\hat C}_{x_i}{\bf g}^{*}_{ji}+\kappa{\bf g}^{H}_{jj}{\bf\hat {C}}_{x_j}{\bf g}^{*}_{jj}, \quad \forall j\in\mathcal{D}.\label{chatz1}
\end{align}}
The interference-plus-noise pseudo-variance is
{\small
\begin{align}
\hat C_{w_k}&=
\hat C_{y_k}-{\bf h}^{H}_{kB}{\bf\hat C}_{x_{B_k}}{\bf h}^{*}_{kB},\quad \forall k\in\mathcal{C},\label{chatw1}\\
\hat C_{q_j}&=
\hat C_{z_j}-{\bf g}^{H}_{jl}{\bf\hat C}_{x_{l}}{\bf g}^{*}_{jl},\quad \forall j\in\mathcal{D},
\label{chatq1}
\end{align}}
where, ${\bf\hat C}_{x_{B_k}}={\bf v}_{B_k}
\mathbb{E}\{ d_{B_k} d_{B_k}\}{\bf v}^{T}_{B_k}$.\\
Based on (\ref{Rr1}) and (\ref{Rr2}), we can denote the achievable rate region of the users as the union of all achievable rates under certain power constraint while preserving the property of the covariance matrix (Hermitian positive semi-definite). Thus, the set of all achievable rates in the network is
{\small
\begin{align}
\Ali{\mathcal{R} \triangleq\bigcup_{\substack{{\rm Tr}({\bf C}_{x_j})\leq  P_j,\\ {\rm Tr}({\bf C}_{x_B})\leq  P_B,\\ {\bf\tilde C}_{x_j}\succeq 0,\  \forall j\in\mathcal{D}\\{\bf\tilde C}_{x_{B_k}}\succeq 0,\ \forall k\in\mathcal{C} }}\{{\bf r}| {\bf 0}\leq {\bf r}\leq {\bf \bar{r}}\},}\label{rateRegion}
\end{align}}
where ${\bf {\bar r}}=[R^{'}_1,..., R^{'}_J, R_1,..., R_{K}]^{T}$.\\
The amount of harvested energies at the users per unit time are
{\small
\begin{align}
e_{k} \leq E_k=&{\bf h}_{kB}^{H}{\bf C}_{x_{B}}{\bf h}_{kB} + \sum_{j=1}^{J}{\bf h}^{H}_{kj}{\bf C}_{x_j}{\bf h}_{kj},\ \forall k\in\mathcal{C},\\
e^{'}_{j} \leq E^{'}_j=& {\bf g}_{jB}^{H}{\bf C}_{x_B}{\bf g}_{jB}+\sum_{\substack{i=1\\i\neq j }}^{J}{\bf g}_{ji}^H{\bf C}_{x_i}{\bf g}_{ji}+\kappa{\bf g}_{jj}^{H}{\rm diag}({\bf C}_{x_j}){\bf g}_{jj},\ \forall j\in\mathcal{D},
\end{align}}
where $E_k$ and $E^{'}_j$ are the incident signal energies at the $k$th cellular user and $j$th D2D user, respectively. The loaded energy is less than these amounts which are denoted by $e_k$ and $e^{'}_j$.\\
Besides rate region, we define the rate-energy region of the $j$th D2D user as
{\small
\begin{align}
\Ali{\mathcal{F}_j \triangleq\bigcup_{\substack{{\rm Tr}({\bf C}_{x_j})\leq  P_j,\\ {\rm Tr}({\bf C}_{x_B})\leq  P_B,\\ {\bf\tilde C}_{x_j}\succeq 0,\  \forall j\in\mathcal{D}\\{\bf\tilde C}_{x_{B_k}}\succeq 0,\ \forall k\in\mathcal{C}\\e^{'}_j\leq E^{'}_j}}\{{\bf f}_j| {\bf 0}\leq {\bf f}_j\leq {\bf \bar{f}}_j\},}
\end{align}}
where ${\bf { f}}_j=[r^{'}_j, e^{'}_j]^T$ is an achievable rate-energy tuple and ${\bf {\bar f}}_j=[R^{'}_j, E^{'}_j]^T$ is the achievable upper-bound.\\
By defining the rate region and rate-energy region of the users, we will discuss the problems in the next section.   
\section{Optimization Problems}
In what follows we present an overview of the considered optimization problems. In section 
\begin{enumerate}[A.]
\item the optimal operating rates of the cellular users while fulfilling energy constraints is investigated.
\item we optimize the operating rate tuples of the full-duplex D2D users given rate demands for the cellular users.
\item the optimal operating rate-energy pairs of D2D users under cellular users' rate constraints are delivered.
\item the optimization problem  considers operating rates and energies of the network jointly under transmit power constraints.
\end{enumerate}

\subsection{Broadcast Users' Rate Region under EH Constraint}
\begin{figure}[t]
\centering
\tikzset{every picture/.style={scale=0.6}, every node/.style={scale=.91}}%
\begin{tikzpicture}

\draw[->,thick] (0,0)--(0,6);
\draw[->,thick] (0,0)--(10,0);

\draw [red] plot [smooth, tension=0.5] coordinates {(0,5) (2,4) (5,3.25) (6,2) (8,1.5) (9.5,0)};

\draw[->,color=blue,dashed] (0,0)--(8.5,0.86);
\draw[->,color=blue,dashed] (0,0)--(1.5,4.07);
\draw [->,color=black,thick] (1.6,0.1) arc (0:90:1.5cm);
\draw[fill=black] (9.5,0) circle (0.1cm);
\draw[fill=black] (8.67,0.9) circle (0.1cm);
\node[rotate=10,color=blue] at (4.5,0.7) {$\Lambda^{*} {\boldsymbol \alpha}$};
\node[rotate=70,color=blue] at (0.8,3) {$\Lambda^{*} {\boldsymbol \alpha}$};
\draw[fill=black] (1.53,4.2) circle (0.1cm);
\draw[fill=black] (0,5) circle (0.1cm);
\node at (5,-0.5) {$R_1$};
\node[rotate=90] at (-0.5,3) {$R_2$};

\draw [fill=black] (8.15,1.4) circle (0.1cm);
\draw [fill=black] (7,1.75) circle (0.1cm);
\draw [fill=black] (5.8,2.15) circle (0.1cm);
\draw [fill=black] (5,3.25) circle (0.1cm);
\draw [fill=black] (4,3.55) circle (0.1cm);
\draw [fill=black] (3,3.75) circle (0.1cm);

\draw[color=red] (-2,-1.5)--(-1.5,-1.5);\node at (1,-1.5) {Pareto boundary};
\draw[color=blue,->,dashed] (3.5,-1.5)--(4,-1.5);\node at (6.2,-1.5) {Scaling direc.};
\draw[->,color=black,thick] (8.8,-1.7) arc (0:90:0.5cm);\node at (11.5,-1.5) {Scanning domain};

\end{tikzpicture}
\caption{Pareto boundary of the achievable rate region of any two conflicting rates. The optimal scaling factor, i.e., $\Lambda^{*}$, in the direction of $\boldsymbol \alpha$ is the maximum of $\Lambda$ while fulfilling the constraints. By formulating the weighted max-min optimization problem and scanning the rate region, the whole point on the Pareto boundary are accessible.}
\label{fig:Chebyshev}
\end{figure}
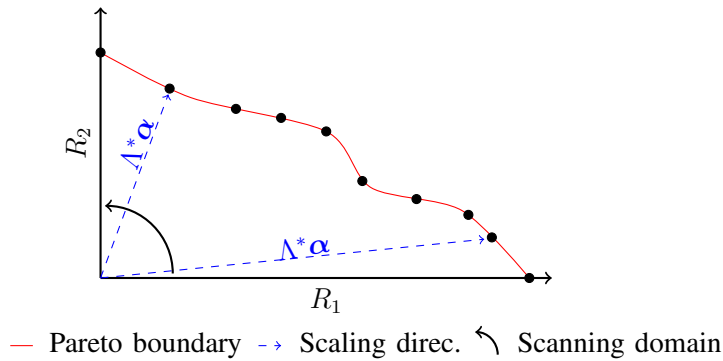

Cellular users are capable of simultaneous ID and EH which is assumed to be achieved by AS receiver structure.  In this section, we study the optimal achievable rate for these users while fulfilling their energy demands. For this, we need to characterize the Pareto boundary of the rate region, on which all the rate pairs are optimal. Here, the Pareto boundary defines the frontier for the achievable rate tuples, such that an increment in the rate of one user inevitably coincides with a decrement in the rate of at least one of the other users. One way to find the Pareto boundary is to maximize sum of the weighted rates,~\cite{Ottersten2014} which is not an efficient way from the complexity perspective. Maximizing the minimum of the weighted rates (known as weighted Chebyshev goal function) is an alternative approach for determining the Pareto boundary, which is shown to be efficient \cite{Ottersten2014}. Here, we focus on the latter. Therefore, the optimization problem that characterizes the Pareto boundary of the achievable rate region is formulated as the weighted max-min problem (weighted Chebyshev problem). In what follows, we formulate this problem under transmit power and harvested energy constraints as
{\small
\begin{subequations}\label{A1}
\begin{align}
\max_{{\bf C}_{x},{\bf\hat {C}}_{x}} \quad &\min_{k\in\mathcal{C}} {\frac{R_k}{\alpha_k}}
\tag{\ref{A1}}\\
\hspace{+0.2cm}{\rm{s.t.}}\ &\Psi_{k}\leq {\bf h}_{kB}^{H}{\bf C}_{x_{B}}{\bf h}_{kB} + \sum_{j=1}^{J}{\bf h}^{H}_{kj}{\bf C}_{x_j}{\bf h}_{kj},\ \forall k\in\mathcal{C},\label{ener1}\\
\hspace{+1cm}&0\leq {\rm{Tr}}({\bf C}_{x_j})\leq P_{x_j},\ \forall j\in\mathcal{D},\label{LimitPower1}\\
\hspace{+1cm}&0\leq {\rm{Tr}}({\bf C}_{x_B})\leq P_{x_B},\label{LimitPower2}\\
\hspace{+1cm}&{\bf\tilde{C}}_{x_j}\succeq 0, \forall j\in\mathcal{D},\label{semiDefin1}\\ 
\hspace{+1cm}&{\bf\tilde{C}}_{x_{B}}\succeq 0,\label{semiDefin2}\\
\hspace{+1cm}&{\rm{rank}}({\bf C}_{x_j})=1,\  \forall j\in\mathcal{D}, \label{rankCons}\\
\hspace{+1cm}&{\rm{rank}}({\bf C}_{x_{B_k}})=1,\ \forall k\in\mathcal{C},\label{rankCons2}
\end{align}
\end{subequations}}
where $\alpha_k$ are the elements of vector ${\boldsymbol \alpha}$, which prioritize the maximization of the minimum of the weighted rates. In other words, ${\boldsymbol \alpha}$ specifies the direction of optimization over the field $\mathbb{R}^{K}$. We define the set $\mathbb{A}$ as, $\mathbb{A}=\{\boldsymbol\alpha\in\mathbb{R}^{K}\given||{\boldsymbol \alpha}||_{1}=1\}$. Solving (\ref{A1}) and scanning the rate region in different directions by means of setting $\bf\alpha\in\{\mathbb{A}\}$ with a predefined resolution will deliver the Pareto-optimal operating points. The set of the Pareto-optimal points specify the Pareto boundary of the rate region. Note that, the convex-hull of all achievable Pareto rate tuples describes the achievable rate region defined in (\ref{rateRegion}).\\
In problem (\ref{A1}), the transmission power is limited by constraints (\ref{LimitPower1}), (\ref{LimitPower2}). On the other hand, the cellular users are required to harvest at least $ \Psi_k$ energy from the received RF signal for full functionality. The energy that has to be obtained by user $k$ is represented in (\ref{ener1}) which needs to be provided by the BS and D2D users. The constraints (\ref{rankCons}) and (\ref{rankCons2}) are due to the feasibility of beamforming vector reconstruction from the optimum covariance matrices, ${\bf C}_{x_{j}}$ and ${\bf C}_{x_{B_k}}$, i.e., feasible beamforming vectors can only be reconstructed from any matrix in the set of rank-1 positive semi-definite matrices.\\
Note that, the optimization parameters are ${\bf C}_{x_{j}},{\bf\hat C}_{x_j} \ \forall j\in\mathcal{D}$ and ${\bf C}_{x_{B_k}},{\bf\hat C}_{x_{B_k}} \ \forall k\in\mathcal{C}$, however we refer to all of them as ${\bf C}_{x}$ and ${\bf\hat C}_{x}$ as the arguments of the objective functions.
\begin{remark}
The energy requirement $\Psi_k,\ \forall k\in\mathcal{C}$ might exceed the BS capability and should be provided to the cellular users by the D2D users. However, this turns the system to a broadcast interference channel. Thus, on one hand, the energy constraint would be fulfilled, and on the other hand the achievable rate would demolish.
\end{remark}
Apparently (\ref{A1}) is a non-convex problem. This can be verified by plugging the entities in (\ref{cy1})-(\ref{cq1}) and (\ref{chaty1})-(\ref{chatq1}) into (\ref{Rr1}) and (\ref{Rr2}). Then we observe that, the objective function is neither a convex nor a concave function with respect to the optimization parameters, i.e. ${\bf C}_{x_{j}}$, ${\bf\hat C}_{x_{j}}$, ${\bf C}_{x_{B_k}}$ and ${\bf\hat C}_{x_{B_k}}$.
\begin{remark}
The objective function is non-convex even in case of proper Gaussian signaling where the achievable rates are bounded by ${R_k^{\mathsf{proper}}}$ and ${R_j^{'\mathsf{proper}}}$ in (\ref{Rr1}) and (\ref{Rr2}). In this case the objective function is the difference of concave functions which is not necessarily convex or concave.
\end{remark}
Problem (\ref{A1}) suffers from non-convexity in the constraint set as well. This is due to the rank-1 constraints (\ref{rankCons}) and (\ref{rankCons2}). Thus, the optimization problem (\ref{A1}) can not be solved except by exhaustive search over the feasible set. However, the computational complexity of exhaustive search is high due to the dimensions of the optimization variables, i.e.,  $M\times M$ and $N\times N$ complex matrices.
Defining, $\Lambda=\min_{k\in\mathcal{C}} \left(\frac{R_k}{\alpha_k}\right)$, the problem is reformulated as
{\small
\begin{subequations}\label{A21}
\begin{align}
&\max_{\Lambda,\mathbf{C}_x,\mathbf{\hat{C}}_{x}} \Lambda\quad {\rm{s.t.}}\ \Lambda\leq \frac{R_k}{\alpha_k},\ \forall k\in\mathcal{C},\,\rm(\ref{ener1})-(\ref{rankCons2}),\tag{\ref{A21}}
\end{align}
\end{subequations}}
where the objective function is translated into the constraint set in the expense of adding an extra scalar parameter. The auxiliary scalar variable $\Lambda$ is maximized in the direction of $\boldsymbol {\alpha}$ in order to get the Pareto-optimal operating point in that direction. This is illustrated in Fig.~\ref{fig:Chebyshev}.

Using the rate expressions in \eqref{Rr1} we have
{\small
\begin{subequations}\label{A22}
\begin{align}
&\max_{\Lambda,\mathbf{C}_x,\mathbf{\hat{C}}_{x}} \Lambda \tag{\ref{A22}}\quad{\rm{s.t.}}\\
&\hspace{0.3cm}\Lambda\leq\frac{1}{\alpha_k} \left(\log\left(\frac{C_{y_k}}{C_{w_k}}\right)+\frac{1}{2}\log\left(\frac{1-C^{-2}_{y_k}|\hat C_{y_k}|^{2}}{1-C^{-2}_{w_k}|\hat C_{w_k}|^{2}}\right)\right), \ \forall k\in\mathcal{C},\label{rarate1}\\
&\hspace{0.5cm}\rm(\ref{ener1})-(\ref{rankCons2}),\nonumber
\end{align}
\end{subequations}}
where the constraint (\ref{rarate1}) consists of the transmit covariance and pseudo-covariance matrices (optimization parameters), which is evident by plugging (\ref{cy1}), (\ref{cw1}), (\ref{chaty1}) and (\ref{chatw1}) into (\ref{Rr1}).
Optimization problem (\ref{A22}) consists of a linear objective function with convex and non-convex constraints.\\
In order to make the problem solvable with less complexity, we proceed with the following separate optimization method:\\
\begin{enumerate}[a)]
\item In the first step, we decouple the optimization problem  (\ref{A22}) into two optimization problems. The first problem contains the first term in the rate expression in constraint (\ref{rarate1}), therefore the optimization variables would only be the transmit covariance matrices. In the second step, we rewrite the problem as a semi-definite program and solve it numerically by interior point methods, \cite{Boyd2004}.\\

\item In the first step, The solutions of (a), i.e. covariance matrices obtained from (a), are used in the second optimization problem which involves the second term of constraint (\ref{rarate1}). Note that the only optimization parameters in the second problem are the transmit pseudo-covariance matrices. In the second step, after some definitions we rewrite the problem as a semi-definite program and solve it numerically by interior point methods.
\end{enumerate}
In the following we discuss the steps in details.
\paragraph{Optimization of Covariance Matrix} 
{\bf Step 1}: First we focus on the first term in the rate expression in (\ref{Rr1}) and (\ref{Rr2}) to optimize the covariance matrices individually. Thus, assuming ${R^{\mathsf{improper}}_k}=0$ and ${R^{'{\mathsf{improper}}}_j}=0$, we replace $C_{y_k}$,  $C_{w_k}$, $C_{z_j}$ and $C_{q_j}$ with the corresponding expressions in (\ref{cy1})-(\ref{cq1}). Consequently, problem (\ref{A1}) simplifies to
{\small
\begin{subequations}\label{A3y}
\begin{align}
\max_{{\bf C}_{x}}\quad &\min_{k\in\mathcal{C}} {\frac{R^{\mathsf{proper}}_k}{\alpha_k}}
\tag{\ref{A3y}}\\
\hspace{0.1cm} {\rm{s.t.}},\ & \Psi_{k}\leq {\bf h}^{H}_{kB}{\bf C}_{x_{B}}{\bf h}_{kB} + \sum_{j=1}^{J}{\bf h}^{H}_{kj}{\bf C}_{x_j}{\bf h}_{kj},\ \forall k\in\mathcal{C},\\
\hspace{1.2cm}&{\bf C}_{x_j}\succeq 0,\ \forall j\in\mathcal{D},\quad {\bf C}_{x_{B}}\succeq 0,\\ 
\hspace{1.2cm}&\rm{(\ref{LimitPower1})-(\ref{LimitPower2}),\ (\ref{rankCons}),\ (\ref{rankCons2}).}\nonumber
\end{align}
\end{subequations}}
By defining $\Gamma = \min_{k\in\mathcal{C}} \left({\frac{R^{\mathsf{proper}}_k}{\alpha_k}}\right)$, we rewrite the problem as
{\small
\begin{subequations}\label{A3}
\begin{align}
\max_{\Gamma,\mathbf {C}_{x}}\quad &\Gamma \tag{\ref{A3}}\\
\hspace{0.1cm}{\rm{s.t.}}\ & \Gamma\leq\Gamma_k^{(1)}\left({\bf C}_{x_{B_k}},{\bf C}_{x_{j}}\right), \ \forall k\in\mathcal{C},\label{r1}\\
\hspace{0.5cm}\ & \Psi_{k}\leq {\bf h}^{H}_{kB}{\bf C}_{x_{B}}{\bf h}_{kB} + \sum_{j=1}^{J}{\bf h}^{H}_{kj}{\bf C}_{x_j}{\bf h}_{kj},\ \forall k\in\mathcal{C},\label{r111}\\
\hspace{0.6cm} &{\bf C}_{x_j}\succeq 0,\ \forall j\in\mathcal{D},\quad {\bf C}_{x_{B}}\succeq 0,\\ 
\hspace{0.6cm}& \rm{(\ref{LimitPower1})-(\ref{LimitPower2}),\ (\ref{rankCons}),\ (\ref{rankCons2}).}\nonumber
\end{align}
\end{subequations}}
where the BS transmit covariance matrix for a particular user, say user $k$, is denoted by ${\bf C}_{x_{B_k}}$. We define $\Gamma_k^{(1)}$ as a function of transmit covariance matrices, i.e.,  $\Gamma_k^{(1)}\left({\bf C}_{x_{B_k}},{\bf C}_{x_{j}}\right)$ as
{\small
\begin{align}
\Gamma_k^{(1)}\left({\bf C}_{x_{B_k}},{\bf C}_{x_{j}}\right)=&\frac{1}{\alpha_k} \log\left(1+\frac{{\bf h}^{H}_{kB}{\bf C}_{x_{B_k}}{\bf h}_{iB}}{\sum_{m=1,m\neq k}^{K}{\bf h}^{H}_{kB}{\bf C}_{x_{B_m}}{\bf h}_{kB}+\sum_{j=1}^{J}{\bf h}^{H}_{kj}{\bf C}_{x_{j}}{\bf h}_{kj}+\sigma^{2}_w+\sigma _n^{2}}\right),\label{rrr1}
\end{align}}

{\bf Step 2}: Now,  we have the separate optimization problem which only depends on the transmit signal covariance matrices. Now, we apply trace operation to~(\ref{r111}) and the numerator and denominator of the expression inside the logarithm in (\ref{rrr1}). By using the shift property of trace and defining ${\bf H}_{ij} = {\bf h}_{ij} {\bf h}^{H}_{ij}$, the optimization problem reduces to
{\small
\begin{subequations}\label{A4}
\begin{align}
\max_{\Gamma,\mathbf{C}_x}\quad &\Gamma \tag{\ref{A4}}\\
\hspace{0cm}{\rm{s.t.}}\ & \Gamma\leq\Gamma_k^{(2)}\left({\bf C}_{x_{B_k}},{\bf C}_{x_{j}}\right) \, \forall k\in\mathcal{C}, \label{e1}\\
\hspace{0.5cm}& \Psi_{k}\leq {\rm Tr}({\bf H}_{kB}{\bf C}_{x_B}) + \sum_{j=1}^{J}{\rm Tr}({\bf H}_{kj}{\bf C}_{x_j}),\ \forall k\in\mathcal{C}, \label{e2}\\
\hspace{0.6cm}& {\bf C}_{x_j}\succeq 0,\ \forall j\in\mathcal{D},\quad  {\bf C}_{x_{B}}\succeq 0,\label{e4}\\ 
\hspace{0.6cm}& \rm{(\ref{LimitPower1})-(\ref{LimitPower2}),\ (\ref{rankCons}),\ (\ref{rankCons2}),}\nonumber
\end{align}
\end{subequations}}
where, $\Gamma_k^{(2)}$ is given as
{\small
\begin{align}
\Gamma_k^{(2)}\left({\bf C}_{x_{B_k}},{\bf C}_{x_{j}}\right)=&\frac{1}{\alpha_k} \log\left(1+\frac{{\rm Tr}({\bf H}_{kB}{\bf C}_{x_{B_k}})}{\sum_{m=1,m\neq k}^{K}{\rm Tr}({\bf H}_{kB}{\bf C}_{x_{B_m}})+\sum_{j=1}^{J}{\rm Tr}({\bf H}_{kj}{\bf C}_{x_{j}})+\sigma^{2}_w+\sigma _n^{2}}\right).\label{rrr2}
\end{align}}
By dropping the rank-1 constraints, i.e., (\ref{rankCons}), (\ref{rankCons2}), problem (\ref{A4}) becomes a convex semi-definite program (SDP) for given $\Gamma$, since the constraint set is convex. In order to get the optimal $\Gamma$ that makes the constraint set feasible, we utilize bisection method.
Therefore, optimization problem (\ref{A4}) can be solved efficiently by checking the feasibility of the constraint set for a given $\Gamma$. Thus, we solve the following feasibility problem for a given $\Gamma$,
{\small
\begin{align}
&\hspace{-0.3cm}{\rm {find}}\quad {\bf C}_{x_j}\in\mathbb{S}^{M}  {\rm and}\ {\bf C}_{x_{B_k}} \in\mathbb{S}^{N},\, \forall j\in\mathcal{D}\ {\rm and}\ \forall k\in\mathcal{C} \label{A5}\\
&\hspace{+.5cm}\rm{s.t.}\ \hspace{+.5cm} (\ref{e1})-(\ref{e4}),\ (\ref{LimitPower1}),(\ref{LimitPower2}), \nonumber
\end{align}}
where $\mathbb{S}^{M}$ and $\mathbb{S}^{N}$ are the cone of $M\times M$ and $N\times N$ Hermitian positive semi-definite matrices, respectively. The solution of problem (\ref{A4}) coincides with the solution of (\ref{A5}) for the maximum $\Gamma$ that makes the constraint set non-empty when the rank-1 constraints are relaxed. In the rest of the paper we denote the optimal covariance matrices of (\ref{A4}) by ${\bf C}^{\star}_{x_j},\ \forall j\in\mathcal{D}$ and $ {\bf C}^{\star}_{x_{B_k}},\ \forall k\in\mathcal{C}$, and the solution of problem (\ref{A4}) by $\Gamma^{\star}$.\\
Based on the solutions of problem (\ref{A4}), we proceed with the two following feasible solutions,
\begin{enumerate} [I.]
\item {\it\bf {The solutions are intrinsically rank-1}}: Then the corresponding rates are achievable, i.e. all the points on the Pareto boundary can be achieved by linear beamforming~\cite{Boyd2004}. Thus, an eigenvalue decomposition of a particular optimal solution, say ${\bf C}^{\star}_{x_j}$ yields,
{\small
\begin{align}
{\bf C}^{\star}_{x_j}= {\bf u}_{j}\beta_k{\bf u}^{H}_{j}={\bf u}_{j}\beta_j^{\frac{1}{2}}\beta_j^{\frac{1}{2}}{\bf  u}^{H}_{j}={\bf t}_j{\bf t}_j^{H},\label{BfVectors}
\end{align}}
where ${\bf u}_{j}$ is the eigenvector corresponding to the single eigenvalue $\beta_j$. Notice that, the beamforming vector for the $j$th D2D user is represented by ${\bf t}_j$.\\ 
\item {\it\bf {The solutions have higher ranks}}: 
We utilize Gaussian randomization procedure,~\cite{Luo2010}, which delivers sub-optimal rank-1 solutions. Gaussian randomization starts by generating finite number of vectors from the Gaussian distribution with zero mean and ${\bf C}^{\star}_{x_j}$ covariance matrix, i.e. $\mathcal{N}\sim({\bf 0},{\bf C}^{\star}_{x_j})$. Then, out of the feasible beamforming solutions, the optimal one which satisfies the constraint set is chosen. Gaussian randomization provides a sub-optimal solution and the quality of the sub-optimality depends on the number of randomizations.\\
\end{enumerate}

\paragraph{Optimization of Pseudo-covariance matrix} 

{\bf Step 1}: By considering the optimal covariance matrix of problem (\ref{A5}), we have the optimal value for the first term in the rate expression in (\ref{Rr1}) which is denoted by $\Gamma^{\star}$. By plugging $\Gamma^{\star}$ into the first term of (\ref{Rr1}), we optimize the pseudo-covariance matrices. Thus, the optimization problem is written as
{\small
\begin{subequations}\label{A6}
\begin{align}
\max_{\Lambda,\mathbf{\hat{C}}_{x}}\quad &\Lambda \tag{\ref{A6}}\\
\hspace{-0.05cm}{\rm{s.t.}}\ &\Lambda\leq \Gamma^{\star}+\frac{1}{2\alpha_k}\log\left(\frac{1-C^{{*}^{-2}}_{y_k}|\hat C_{y_k}|^{2}}{1-C^{{*}^{-2}}_{w_k}|\hat C_{w_k}|^{2}}\right),\ \forall k\in\mathcal{C},\label{r3}\\
\hspace{0.6cm}&{\bf\tilde{C}}_{x_j}\succeq 0,\ \forall j\in\mathcal{D},\quad {\bf\tilde{C}}_{x_B}\succeq 0\label{Pseudo39B}
\end{align}
\end{subequations}}
where the power and energy constraints are dropped since they are embedded in the covariance part of the augmented covariance matrix.\\
{\bf Step 2}: The optimization problem (\ref{A6}) is solved efficiently in the Appendix.
\subsection{Rate Region of the D2D users}
The coexistence of D2D communication in the crowd of cellular users requires the study of the achievable rate region of the full-duplex D2D users while guaranteeing rate demands of the other users. We can resemble this case as a network with cognitive users, where the cellular users are the primary users and the D2D users are the secondary users with cognition. Particularly, in an underlay cognitive network where D2D users are active only in case of fulfilling the primary users' demands. In this section we assume the case that the primary users request only information and we formulate the maximum achievable rate-tuples for the D2D users. The problem is written as
{\small
\begin{subequations}\label{A711}
\begin{align}
\max_{{\bf C}_{x},{\bf\hat {C}}_{x}} \quad &\min_{j\in\mathcal{D}} {\frac{R^{'}_j}{\alpha_j}}
\tag{\ref{A711}}\\
\hspace{0.3cm}{\rm{s.t.}}\ &\Sigma_k\leq\ \Sigma_k^{(1)}\left({\bf C}_{x_{B_k}},{\bf C}_{x_{j}}\right), \ \forall k\in\mathcal{C},\\
\hspace{0.9cm}&{\rm(\ref{LimitPower1})-(\ref{rankCons2})},\nonumber
\end{align}
\end{subequations}}
where, $R^{'}_j$ is the achievable rate for the $j$th full-duplex D2D user that is given in (\ref{Rr2}) and $\Sigma_k$ is the rate demand for $k$th cellular user. Note that, $\Sigma_k^{(1)}$ is given by
{\small
\begin{align}
\Sigma_k^{(1)}\left({\bf C}_{x_{B_k}},{\bf C}_{x_{j}}\right)=\log\left(1+\frac{{\bf h}^{H}_{kB}{\bf C}_{x_{B_k}}{\bf h}_{kB}}{\sum_{m=1,m\neq k}^{K}{\bf h}^{H}_{kB}{\bf C}_{x_{B_m}}{\bf h}_{kB}+\sum_{j=1}^{J}{\bf h}^{H}_{kj}{\bf C}_{x_{j}}{\bf h}_{kj}+\sigma^{2}_w+\sigma _n^{2}}\right).\label{rrr3}
\end{align}}
Hence, the objective functions are composed of the covariance and pseudo-covariance matrices of the transmit signals. To solve this problem we proceed with the same procedure as described in the last section. First we optimize the covariance matrix assuming $R^{'\mathsf{improper}}_j=0$,  which is
{\small
\begin{subequations}\label{A712}
\begin{align}
\max_{{\bf C}_{x}} \quad &\min_{j\in\mathcal{D}} {\frac{R^{'\mathsf{proper}}_j}{\alpha_j}}
\tag{\ref{A712}}\\
\hspace{0.3cm}&{\rm{s.t.}}\ \Sigma_k\leq\ \Sigma_k^{(1)}\left({\bf C}_{x_{B_k}},{\bf C}_{x_{j}}\right), \ \forall k\in\mathcal{C},\\
\hspace{0.9cm}&{\rm(\ref{LimitPower1})-(\ref{rankCons2})}.\nonumber
\end{align}
\end{subequations}}
By defining, $\Gamma=\min_{j\in\mathcal{D}} {\frac{R^{'\mathsf{proper}}_j}{\alpha_j}}$ we formulate the respective SDP problem as
{\small
\begin{subequations}\label{A7}
\begin{align}
\max_{\Gamma,\mathbf{C}_x}\quad &\Gamma \tag{\ref{A7}}\\
\hspace{0.4cm}{\rm{s.t.}}\ &\Gamma\leq\Gamma_j^{(3)}\left({\bf C}_{x_{B_k}},{\bf C}_{x_{j}}\right), \, \forall j\in\mathcal{D},\\
\hspace{1cm}&\Sigma_k\leq\ \Sigma_k^{(2)}\left({\bf C}_{x_{B_k}},{\bf C}_{x_{j}}\right), \, \forall k\in\mathcal{C},\\
\hspace{1cm} &\rm(\ref{LimitPower1})-(\ref{rankCons2}),\nonumber
\end{align}
\end{subequations}}
where $\Gamma_j^{(3)}$ and $\Sigma_k^{(2)}$ are defined as
{\small
\begin{align}
\Gamma_j^{(3)}\left({\bf C}_{x_{B_k}},{\bf C}_{x_{j}}\right)=&\frac{1}{\alpha_j} \log\left(1+\frac{{\rm Tr}({\bf G}_{jl}{\bf C}_{x_{l}})}{{\rm Tr}({\bf G}_{jB}{\bf C}_{x_{B}})+\sum_{\substack{i=1\\i\neq j,l}}^{J}{\rm Tr}({\bf H}_{ji}{\bf C}_{x_{i}})+\kappa {\rm Tr}\left({\bf G}_{jj}{\rm diag}({\bf C}_{x_{j}})\right)+\sigma^{2}_w+\sigma _n^{2}}\right),\label{rrr4}\\
\Sigma_k^{(2)}\left({\bf C}_{x_{B_k}},{\bf C}_{x_{j}}\right)=&\log\left(1+\frac{{\rm Tr}({\bf H}_{kB}{\bf C}_{x_{B_k}})}{\sum_{m=1,m\neq k}^{K}{\rm Tr}({\bf H}_{kB}{\bf C}_{x_{B_m}})+\sum_{j=1}^{J}{\rm Tr}({\bf H}_{kj}{\bf C}_{x_{j}})+\sigma^{2}_w+\sigma _n^{2}}\right),\label{rrr5}
\end{align}}
respectively.
By ignoring the rank-1 constraints, we solve the SDP efficiently. Furthermore, we compensate the relaxation by Gaussian randomization method in order to get a feasible optimal solution. Note that the optimization problem of (\ref{A7}) yields the optimal transmit covariance matrices while the rate region can be further improved by optimization over the pseudo-covariance matrices. Optimizing pseudo-covariance matrices for this problem is similar to problem (\ref{A22}) which can be solved similarly as in Appendix.
\subsection{Joint Rate-Energy Optimization (full-duplex D2D users)}
In this subsection we present the rate-energy region of the D2D users assuming self-interference and transmitter noise with active base station. The full-duplex D2D users are equipped with a single receive antenna. In a single-antenna receiver, either information out of the received signal can be extracted or the energy unless by power splitting (PS) or time sharing (TS). 
First, we study the PS receiver structure, where each D2D user splits the received signals power and decodes the information of one portion and captures the energy from the other portion.
We formulate the optimization problem that achieves the Pareto boundary of the rate-energy region as
{\small
\begin{subequations}\label{B1}
\begin{align}
\max_{{\bf C}_{x},{\bf\hat {C}}_{x}} \quad &\min \left(\frac{R^{'}_j}{\alpha_1}, \frac{E^{'}_j}{\alpha_2}\right) \tag{\ref{B1}}\\
\hspace{-3cm}&{\rm{s.t.}}\ \Sigma_k\leq\ \Sigma_k^{(2)}\left({\bf C}_{x_{B_k}},{\bf C}_{x_{j}}\right),\, \forall k\in\mathcal{C}\label{ERTDMA11},\\
\hspace{-2.5cm}\ &\rm(\ref{LimitPower1})-(\ref{rankCons2}),\nonumber
\end{align}
\end{subequations}}
where, $0\leq\alpha_1\leq 1$ and $\alpha_2=1-\alpha_1$.\\
We define $\eta$ as the power splitting factor, so that $\eta=1$ corresponds with pure information detection and $\eta=0$ is associated with pure energy harvesting. Thus, simultaneous EH and ID occurs by setting $0<\eta< 1$.   
By this definition, we first optimize the covariance matrices as,
{\small
\begin{subequations}\label{B11}
\begin{align}
\max_{{\bf C}_{x},\eta}\quad &\min \left(\frac{R^{'\mathsf{proper}}_j(\eta)}{\alpha_1}, \frac{E^{'}_j(\eta)}{\alpha_2}\right) \tag{\ref{B11}}\\
\hspace{-4.5cm}&{\rm{s.t.}}\ \Sigma_k\leq\ \Sigma_k^{(2)}\left({\bf C}_{x_{B_k}},{\bf C}_{x_{j}}\right),\, \forall k\in\mathcal{C},\label{ERTDMA112}\\
\hspace{-4cm}\ &\rm(\ref{LimitPower1})-(\ref{rankCons2}),\nonumber
\end{align}
\end{subequations}}
where $R^{'\mathsf{proper}}_j(\eta)$ and $E^{'}_j(\eta)$ are the achievable rates and energies at the $j$th full-duplex D2D user, respectively. The achievable energy is formulated as
{\small
\begin{align}
 E^{'}_j(\eta)=&(1-\eta)({\rm Tr}({\bf G}_{ji}{\bf C}_{x_{i}})+{\rm Tr}({\bf G}_{jB}{\bf C}_{x_{B}})+\kappa {\rm Tr}({\bf G}_{jj}{\rm diag}({\bf C}_{x_{j}}))),
\end{align}}
and the achievable rate is expressed as
{\small
\begin{align}
R^{'\mathsf{proper}}_j(\eta)&=\log\left
(1+\frac{\eta{\rm Tr}({\bf G}_{jl}{\bf C}_{x_{l}})}{\eta\left({\rm Tr}({\bf G}_{jB}{\bf C}_{x_{B}})+\sum_{\substack{i=1\\i\neq j,l}}^{J}{\rm Tr}({\bf H}_{ji}{\bf C}_{x_{i}})+\kappa {\rm Tr}\left({\bf G}_{jj}{\rm diag}({\bf C}_{x_{j}})\right)+\sigma^{2}_w\right)+\sigma _n^{2}}\right),\label{rrr6}
\end{align}}
Note that the achievable rates and energies are functions of the power splitting coefficient, $\eta$. We define, $\Gamma=~\min \left(\frac{R^{'\mathsf{proper}}_j(\eta)}{\alpha_1}, \frac{E^{'}_j(\eta)}{\alpha_2}\right)$.
Then problem (\ref{B11}) is rewritten as
{\small
\begin{subequations}\label{B111}
\begin{align}
\max_{{\bf C}_{x},\eta}\quad &\Gamma \tag{\ref{B111}}\\
\hspace{-0.8cm}{\rm{s.t.}}\ &\Gamma\leq \frac{R^{'\mathsf{proper}}_j(\eta)}{\alpha_1}, \, \forall j\in\mathcal{D},\\
\hspace{-0.2cm}&\Gamma\leq\frac{E^{'}_j(\eta)}{\alpha_2}, \, \forall j\in\mathcal{D},\\
\hspace{-0.2cm} &\Sigma_k\leq\Sigma_k^{(2)}\left({\bf C}_{x_{B_k}},{\bf C}_{x_{j}}\right), \, \forall k\in\mathcal{C},\label{ERTDMA113}\\
\hspace{-0.3cm}\ &\rm(\ref{LimitPower1})-(\ref{rankCons2}).\nonumber
\end{align}
\end{subequations}}
By exhaustive search over $\eta$ and bisection over $\Gamma$, the feasibility check problem can be efficiently solved (the feasibility check problem for (\ref{B111}) can be formulated in a similar way as Problem (\ref{A1})). If the optimal solutions do not fulfill the rank-1 constraints, the Gaussian randomization procedure finds a sub-optimal solution correspondingly. The pseudo-covariance matrices are optimized by some  vector definitions similar to the procedure elaborated in the Appendix.\\
Time sharing is the other strategy that could be utilized for joint ID and EH in a single antenna receivers. The achievable rate-energy region for TS receivers can be found by determining the two extremum points which are achievable by pure ID and pure EH.\\
Power splitting and TS receivers characterize the trade-off between energy and rate of a particular user ($R^{'}_1$-$E^{'}_1$ or $R^{'}_2$-$E^{'}_2$). By setting one user to purely decode information and the other user to purely harvest energy, we can study the trade-off between the objectives of different users. Suppose one user, say user 1, harvests energy while the other user, say user 2, detects information and vice versa. Therefore, we are interested in the rate-energy region ($R^{'}_2$-$E^{'}_1$ or $R^{'}_1$-$E^{'}_2$) while guaranteeing cellular users' demands. This is achieved by scanning the rate-energy region in the positive quadrant of $\mathbb{R}^{2}$.\\
In this case, the problem is expressed as
{\small
\begin{subequations}\label{B112}
\begin{align}
\max_{{\bf C}_{x},{\bf\hat {C}}_{x},\ j\neq i}\quad & \min \left(\frac{R^{'}_i}{\alpha_1}, \frac{E^{'}_j}{\alpha_2}\right) \tag{\ref{B112}}\\
{\rm{s.t.}}\ &\Sigma_k\leq\ \Sigma_k^{(2)}\left({\bf C}_{x_{B_k}},{\bf C}_{x_{j}}\right), \, \forall k\in\mathcal{C},\label{ERTDMA114}\\
&\rm(\ref{LimitPower1})-(\ref{rankCons2}),\nonumber
\end{align}
\end{subequations}}
It is important to note that, not only the optimum covariance and pseudo-covariance but also the optimum rate-energy pair is crucial, so that one could decide which user to detect information and which user to harvest energy. This problem is solved similarly and we skip reformulations.
\subsection{Joint Rate-Energy Optimization Simultaneously}
Simultaneous optimization of the rates and the energies jointly might be considered if the nodes are capable of EH and ID at the same time. We can think of this by implementing one extra antenna at the receivers. Thus, one antenna is used for information detection, while the other harvests energy, (AS). In this paper we do not discuss the optimality of using both antennas at the receivers for improving achievable rates of the users. Thus, we stick to a single-antenna information reception system and an auxiliary antenna for energy harvesting. The problem delivers the ($2K+J$)-dimensional rate-energy region, where $K$ cellular users simultaneously harvest energy and decode information by AS and $J$ full-duplex D2D users are decoding information only. We formulate the joint rate and energy maximization problem as
{\small
\begin{align}
&\hspace{-.5cm}\max_{{\bf C}_{x},{\bf\hat {C}}_{x}} \min_{\phi\in\Phi} \left(\frac{{\Omega}_\phi} {\beta_\phi}\right) \quad {\rm{s.t.}}\, \rm(\ref{LimitPower1})-(\ref{rankCons2}),\label{JointOpt}
\end{align}}
where, $\sum_{\phi=1}^{2K+J} \beta_\phi=1$ and $\Omega_\phi$ is defined as
{\small
\begin{align}
{\Omega_\phi}=\begin{cases}
R^{'}_\phi,\quad &\phi=1,...,J\\
R_{\phi-J},\quad &\phi=J+1,...,K+J\\
E_{\phi-K-J},\quad &\phi=K+J+1,...,2K+J\\
\end{cases}
\end{align}}
This optimization problem is solved using separate optimization method that has already been discussed. According to this optimization problem we can observe all optimal rate-energy pairs of the system.\\
We proceed with the numerical results of the proposed optimization problems in the next section.
\section{Numerical Results}
In this section, we study the simulation results and discuss the insights for the rate and energy optimization problems formulated in the former sections.\\
Proper Gaussian signaling is numerically shown not to be optimal for the investigated setup. Different aspects of the network are discussed and the achievable rate region and rate-energy trade-off are delivered. For simplicity in illustration,
\begin{itemize}
\item We assume two active cellular users in the network,
\item We limit transmit antennas to two.
\end{itemize}
The transmit power budget at the BS and full-duplex nodes as assumed to be $P_\text{B}=4$ and $P_j=2,\ \forall j$, respectively. The AWGN variance is assume to be $\sigma^{2}_n=1$.
\begin{figure*}
\centering
\begin{minipage}[b]{0.47\textwidth}
\subfigure[Rate region improvement of broadcast channel by improper Gaussian signaling.]{
\tikzset{every picture/.style={scale=.85}, every node/.style={scale=.8}}%
\definecolor{mycolor1}{rgb}{0.00000,0.00000,0.56250}%
\definecolor{mycolor2}{rgb}{0.56250,1.00000,0.43750}%
\begin{tikzpicture}

\begin{axis}[%
width=3.5in,
height=2.5in,
scale only axis,
xmin=0,
xmax=1.6,
xmajorgrids,
xlabel={$R_1$ (bits/channel use)},
ymin=0,
ymax=1.6,
ymajorgrids,
ylabel={$R_2$ (bits/channel use)},
ylabel near ticks,
legend style={at={(axis cs: 1.6,1.6)},anchor= north east, draw=black,fill=white,legend cell align=left}
]
\addplot [color=magenta,solid,mark=asterisk,mark options={solid}]
  table[row sep=crcr]{2.8278138597404e-07	1.56559615471742\\
0.206390352804864	1.15729018363456\\
0.341332016883039	0.951766775333639\\
0.456532263430116	0.802796079029235\\
0.566195450886666	0.67876361362845\\
0.678763619355812	0.566195455728119\\
0.802796093329249	0.456532272348685\\
0.951766802145933	0.341332028999964\\
1.15729023736352	0.206390366960951\\
1.56559640207954	2.82772132820241e-07\\
};
\addlegendentry{MRT, proper Gaussian};

\addplot [color=blue,solid]
  table[row sep=crcr]{5.35026467574085e-11	1.5655953060376\\
0.0787194352488501	1.41694465417754\\
0.152002097598464	1.29201689945986\\
0.221809580209113	1.18297063303736\\
0.289397607364586	1.08526264415409\\
0.355680323215227	0.995897033074327\\
0.421357798511632	0.912935641822755\\
0.487033212448894	0.834910217120235\\
0.553255872621222	0.760724722558799\\
0.62056274365548	0.689513976158875\\
0.689513976117802	0.620562743655306\\
0.760724719231662	0.553255875905799\\
0.834910219094289	0.487033211663129\\
0.91293564042994	0.421357799155449\\
0.995897037081506	0.355680320237505\\
1.0852424227874	0.289399502242535\\
1.18297155277218	0.221809336589536\\
1.29201691373365	0.152002087482145\\
1.41694468761641	0.0787194121477638\\
1.56559530021211	1.08062780945772e-10\\
};
\addlegendentry{Optimal proper Gaussian};
\addplot [color=green!50!black,dashed]
  table[row sep=crcr]{5.35026467574085e-11	1.5655953060376\\
0.078886080393924	1.41978071284907\\
0.153082718017134	1.30115869555892\\
0.225037309634702	1.20016759347926\\
0.296402095007774	1.1115294703019\\
0.368539249605212	1.03190200491041\\
0.442683714969745	0.95914179025477\\
0.520143914627546	0.891671409308046\\
0.602420301566881	0.828325803361512\\
0.691339320649601	0.768154497984711\\
0.768167035653366	0.691350560966026\\
0.82834654222468	0.60243549494846\\
0.891693997245708	0.520157087882209\\
0.95914179183978	0.4426837259414\\
1.03190201352653	0.368539250244946\\
1.11150974867042	0.296406201227987\\
1.20016849422274	0.225037010567903\\
1.30115867586929	0.153082641901932\\
1.41978075573577	0.0788857451464867\\
1.56559530021211	1.08062780945772e-10\\
};
\addlegendentry{Optimal improper Gaussian};

\addplot [color=black,mark=*,mark options={solid}]
  table[row sep=crcr]{2.48451259565741e-11	1.56559529592192\\
0.083389621533721	1.5010163493211\\
0.168487638195098	1.43213993065826\\
0.254855243039878	1.35922120202899\\
0.342035944857589	1.28263316620483\\
0.429585080696534	1.20283454503421\\
0.517088264686379	1.12035218996624\\
0.604184967805728	1.03573786054634\\
0.690578387221469	0.949537413744094\\
0.776040252300243	0.862262265682921\\
0.862262265335107	0.776040254919077\\
0.949537410320205	0.690578391998603\\
1.03573787444602	0.604184948187171\\
1.12035221670623	0.517088235169533\\
1.20286138176313	0.429594765198095\\
1.2826331741363	0.342035934128029\\
1.35922119964579	0.254855255394329\\
1.43214004977645	0.168487462976521\\
1.50101631746246	0.0833896569862794\\
1.56559530086937	4.96467311705828e-11\\
};
\addlegendentry{Capacity};
\end{axis}
\end{tikzpicture}%
\label{BcChannelNoEH}
}
\end{minipage}
\quad
\begin{minipage}[b]{0.47\textwidth}
\subfigure[Rate region improvement by improper signaling under EH constraint, $\Psi_{k}=6,\forall k$.]{
\tikzset{every picture/.style={scale=.85}, every node/.style={scale=.8}}%
\definecolor{mycolor1}{rgb}{0.00000,0.00000,0.56250}%
\definecolor{mycolor2}{rgb}{0.56250,1.00000,0.43750}%
\begin{tikzpicture}

\begin{axis}[%
width=3.5in,
height=2.5in,
scale only axis,
xmin=0,
xmax=1.6,
xlabel={$R_1$ (bits/channel use)},
xmajorgrids,
ymin=0,
ymax=1.6,
ylabel={$R_2$ (bits/channel use)},
ylabel near ticks,
ymajorgrids,
legend style={at={(axis cs: 1.6,1.6)},anchor= north east, draw=black,fill=white,legend cell align=left}
]
\addplot [color=blue,solid, mark=o]
  table[row sep=crcr]{5.35026467574085e-11	1.5655953060376\\
0.0787194352488501	1.41694465417754\\
0.152002097598464	1.29201689945986\\
0.221809580209113	1.18297063303736\\
0.289397607364586	1.08526264415409\\
0.355680323215227	0.995897033074327\\
0.421357798511632	0.912935641822755\\
0.487033212448894	0.834910217120235\\
0.553255872621222	0.760724722558799\\
0.62056274365548	0.689513976158875\\
0.689513976117802	0.620562743655306\\
0.760724719231662	0.553255875905799\\
0.834910219094289	0.487033211663129\\
0.91293564042994	0.421357799155449\\
0.995897037081506	0.355680320237505\\
1.0852424227874	0.289399502242535\\
1.18297155277218	0.221809336589536\\
1.29201691373365	0.152002087482145\\
1.41694468761641	0.0787194121477638\\
1.56559530021211	1.08062780945772e-10\\
};
\addlegendentry{Proper Gaussian (no EH constraints)};
\addplot [color=red,dashed, mark=*, ,mark options={solid}]
  table[row sep=crcr]{5.35026467574085e-11	1.5655953060376\\
0.078886080393924	1.41978071284907\\
0.153082718017134	1.30115869555892\\
0.225037309634702	1.20016759347926\\
0.296402095007774	1.1115294703019\\
0.368539249605212	1.03190200491041\\
0.442683714969745	0.95914179025477\\
0.520143914627546	0.891671409308046\\
0.602420301566881	0.828325803361512\\
0.691339320649601	0.768154497984711\\
0.768167035653366	0.691350560966026\\
0.82834654222468	0.60243549494846\\
0.891693997245708	0.520157087882209\\
0.95914179183978	0.4426837259414\\
1.03190201352653	0.368539250244946\\
1.11150974867042	0.296406201227987\\
1.20016849422274	0.225037010567903\\
1.30115867586929	0.153082641901932\\
1.41978075573577	0.0788857451464867\\
1.56559530021211	1.08062780945772e-10\\
};
\addlegendentry{Improper Gaussian (no EH constraints)};
\addplot [color=blue,solid]
  table[row sep=crcr]{2.67652566776633e-11	0.717855599666334\\
0.0777007852073747	0.621605815890539\\
0.153653175944347	0.537787923405181\\
0.230740082320405	0.461476796662074\\
0.311593047883481	0.389490741136839\\
0.399254818547616	0.319403993102968\\
0.497924178299107	0.248963382576732\\
0.614389908351757	0.1755408711239\\
0.761637192126055	0.0952051206095612\\
0.971426780969295	5.35005373336617e-12\\
};
\addlegendentry{Proper Gaussian (with EH constraints)};
\addplot [color=red,dashed]
  table[row sep=crcr]{2.29787560495631e-05	1.09440197311824\\
0.144569617415275	1.04590755695615\\
0.278887824743991	0.976097963917482\\
0.42582901903946	0.851650236273764\\
0.559851526466397	0.699813213818211\\
0.677199006853357	0.541760307346248\\
0.773298219486045	0.394903238469037\\
0.859324366199657	0.305726230538093\\
0.953357666398605	0.160975250775837\\
1.10039996230076	2.70833668825077e-08\\
};
\addlegendentry{Improper Gaussian (with EH constraints)};

\end{axis}
\end{tikzpicture}%
\label{BcChannelWithEH}
}
\end{minipage}

\caption{Achievable rate region of the cellular users.}

\end{figure*}
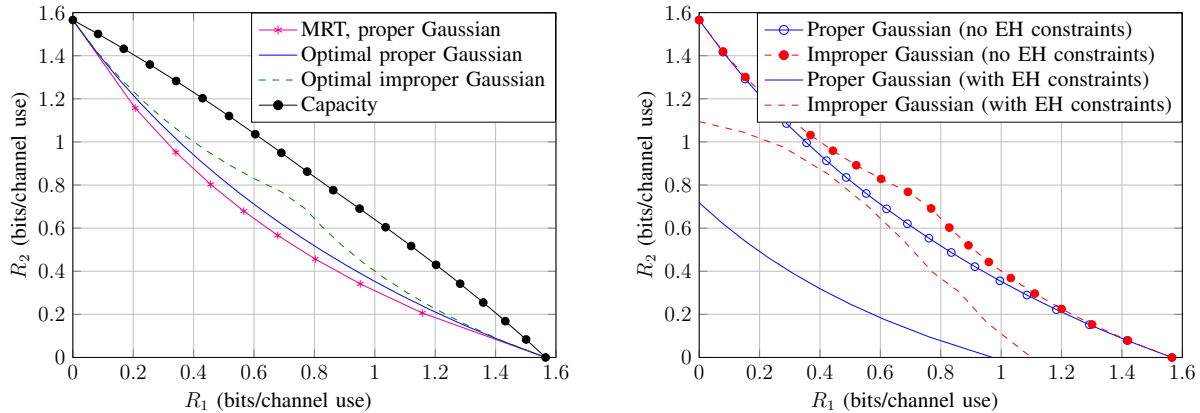

\subsection{Cellular Users' Rate Region}
In this subsection, we discuss the rate region improvement of the cellular users when allowing improper Gaussian signaling. We assume that an extra receive antenna is employed in the cellular users in order to obtain the required amount of energy from the RF signals. The channel that is experienced by the information decoding chain and energy harvesting chain is assumed to be fully correlated. The discussions in this section are based on the solution of problem (\ref{A1}).\\
It is of importance to note that, the capacity of the MIMO Gaussian broadcast channel is achieved by treating interference as noise (TIN) in the receivers and dirty paper coding (DPC) and time-sharing at the transmitter with proper Gaussian signaling,~\cite{Vishwanath2003},~\cite{Vishwanath20032}. In order to show the performance of Gaussian signaling with linear precoding, we compare the achievable rate region with the optimal scheme (DPC which is a non-linear precoding scheme). Figure \ref{BcChannelNoEH} compares the achievable rate region of improper Gaussian signaling, proper Gaussian signaling and DPC scheme. Moreover, we compare the performance of optimized beamforming solution with the solution of maximum ratio transmission (MRT). Notice that, by MRT each user transmits in the direction of its own channel. Hence, the power allocation problem for MRT turns to be a geometric program which can be solved efficiently. Notice that, since MRT is not an optimal solution, some energy demands are not satisfied. Therefore, the problem becomes infeasible by MRT, however it is feasible by the optimized beamforming solution.

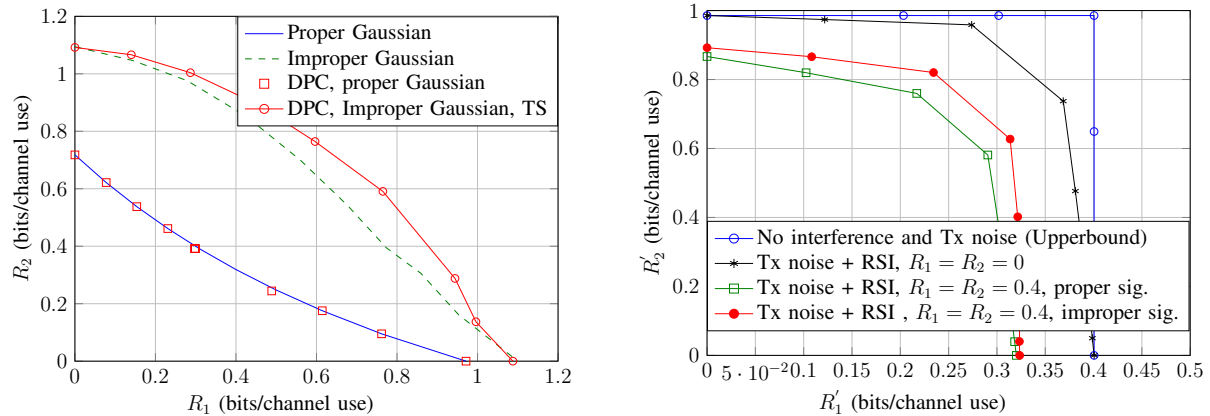
\begin{figure*}

\begin{minipage}[b]{0.47\textwidth}
\subfigure[Antenna separation is the receiver structure for ID and EH purposes. Note that, TS refers to time-sharing between decoding orders.]{
\tikzset{every picture/.style={scale=.85}, every node/.style={scale=.8}}%
\definecolor{mycolor1}{rgb}{0.00000,0.75000,0.75000}%
\definecolor{mycolor2}{rgb}{0.75000,0.00000,0.75000}%
\definecolor{mycolor3}{rgb}{0.75000,0.75000,0.00000}%
\begin{tikzpicture}

\begin{axis}[%
width=3.5in,
height=2.5in,
scale only axis,
xmin=0,
xmax=1.2,
xmajorgrids,
ymin=0,
ymax=1.2,
ymajorgrids,
xlabel={$R_1$ (bits/channel use)},
ylabel={$R_2$ (bits/channel use)},
ylabel near ticks,
legend style={at={(axis cs: 1.2,1.2)},anchor= north east, draw=black,fill=white,legend cell align=left}
]
\addplot [color=blue,solid]
  table[row sep=crcr]{2.67652566776633e-11	0.717855599666334\\
0.0777007852073747	0.621605815890539\\
0.153653175944347	0.537787923405181\\
0.230740082320405	0.461476796662074\\
0.311593047883481	0.389490741136839\\
0.399254818547616	0.319403993102968\\
0.497924178299107	0.248963382576732\\
0.614389908351757	0.1755408711239\\
0.761637192126055	0.0952051206095612\\
0.971426780969295	5.35005373336617e-12\\
};
\addlegendentry{Proper Gaussian};
\addplot [color=green!50!black,dashed]
  table[row sep=crcr]{2.29787560495631e-05	1.09440197311824\\
0.144569617415275	1.04590755695615\\
0.278887824743991	0.976097963917482\\
0.42582901903946	0.851650236273764\\
0.559851526466397	0.699813213818211\\
0.677199006853357	0.541760307346248\\
0.773298219486045	0.394903238469037\\
0.859324366199657	0.305726230538093\\
0.953357666398605	0.160975250775837\\
1.10039996230076	2.70833668825077e-08\\
};
\addlegendentry{Improper Gaussian};

\addplot [color=red,only marks,mark=square,mark options={solid}]
  table[row sep=crcr]{0	0.717855598575138\\
0.0777008647160042	0.621605766632715\\
0.153657537159624	0.537787935875898\\
0.23073948540351	0.461477161139405\\
0.298297908364859	0.391657351559965\\
0.29829450583036	0.391659447722716\\
0.298293372487241	0.391660162270783\\
0.298293372515024	0.391660160028686\\
0.29829790789862	0.391657263024855\\
0.298296774937579	0.391657988672626\\
0.48881256597227	0.244395658209152\\
0.614390359131258	0.175540384697533\\
0.76163728563555	0.0952050178863594\\
0.9714267098629	0\\
};
\addlegendentry{DPC, proper Gaussian};

\addplot [color=red,solid,mark=o,mark options={solid}]
  table[row sep=crcr]{2.32668497468502e-07	1.09184979632052\\
0.14009376831963	1.06632681024314\\
0.286832194225098	1.00388770313359\\
0.448639164319969	0.897274704537564\\
0.596471464019856	0.764373207467243\\
0.764915886046818	0.590815339245402\\
0.944061168149305	0.287912615993188\\
0.996044738330709	0.137114273957985\\
1.08787114129709	6.94083599483974e-08\\
};
\addlegendentry{DPC, Improper Gaussian, TS};

\end{axis}
\end{tikzpicture}%

\label{fig:DPCwithImproperSig2}
}
\end{minipage}
\quad
\begin{minipage}[b]{0.47\textwidth}
\subfigure[Rate region of D2D users under cellular users' rate constraints]{
\tikzset{every picture/.style={scale=.85}, every node/.style={scale=.8}}%
\begin{tikzpicture}

\begin{axis}[%
width=3.5in,
height=2.5in,
scale only axis,
xmin=0,
xmax=0.5,
xmajorgrids,
ymin=0,
ymax=1,
ymajorgrids,
xlabel={$R^{'}_1$ (bits/channel use)},
ylabel={$R^{'}_2$ (bits/channel use)},
ylabel near ticks,
legend style={at={(axis cs: 0,0.07)},anchor= south west, draw=black,fill=white,legend cell align=left}
]
\addplot [color=blue,solid, mark=o]
  table[row sep=crcr]{0	0.985498660947447\\
0.203307421437286	0.985498645666107\\
0.301769618389154	0.985499171433183\\
0.4005379291539     0.985499171433183\\
0.4005379291539	    0.649179369462311\\
0.4005379291539	    0.3\\
0.400537762665856	0\\
};
\addlegendentry{No interference and Tx noise (Upperbound)};

\addplot [color=black,solid,mark=asterisk]
  table[row sep=crcr]{0	0.985498660947447\\
0.121740543341248	0.973932140070829\\
0.273853094752039	0.958491213411813\\
0.368621151715534	0.737237995138384\\
0.381168373073703	0.476459705162627\\
0.388413609134961	0.310728396015555\\
0.393085830754869	0.196540351394029\\
0.396333870058057	0.11323822897779\\
0.39871027560229	0.0498444669148228\\
0.400537564793395	0\\
};
\addlegendentry{Tx noise + RSI, $R_1=R_2=0$};

\addplot [color=green!50!black,solid,mark=square]
  table[row sep=crcr]{0	0.86675261431553\\
0.102459739407342	0.819682195701926\\
0.217047746665604	0.75966655231749\\
0.290541826621958	0.581104806373219\\
0.301764576030689	0.377229894590192\\
0.308452173892566	0.246758368535875\\
0.312866366184147	0.156430977016999\\
0.315988573114908	0.0902815933263418\\
0.318304995312373	0.0398135533658248\\
0.32010182042828	    0\\
};
\addlegendentry{Tx noise + RSI, $R_1=R_2=0.4$,  proper sig.};
\addplot [color=red,solid, mark=*]
  table[row sep=crcr]{0	0.892287243735772\\
0.108262958561294	0.866104790271536\\
0.234369127158955	0.820281298497656\\
0.313645676082461	0.627313251102847\\
0.321510347843622	0.401913363745466\\
0.324908212574019	0.259930150193578\\
0.324667375484538	0.16235476158192\\
0.323469505222323	0.0924347706144331\\
0.323274938277401	0.0404494348563309\\
0.323511205989243	0\\
};
\addlegendentry{Tx noise + RSI , $R_1=R_2=0.4$, improper sig.};
\end{axis}
\end{tikzpicture}%
\label{Fd_D2d_1}
}
\end{minipage}

\caption{Achievable rate region of the cellular and D2D users}

\end{figure*}

\begin{remark}
The broadcast channel investigated in this paper suffers from interference caused by the D2D users which are active in order to satisfy the cellular users' energy demands. Note that, for the Gaussian broadcast channel with EH constraints, the capacity is still unknown.
\end{remark}
Now, it is required that, the cellular users should obtain particular RF energy from the environment. The case might happen that the required energy is far more than that exists in their surroundings. Hence, power should be transmitted to the cellular users in order to fulfill the energy demands. Assuming that the demanded energy is provided by the BS, it is rate-optimal for the cellular users if the D2D users remain silent or do zero-forcing in order not to cause interference at the cellular users. But if the demanded energy is more than the BS capability, the D2D users get activated to fulfill the cellular users' energy demands. In this case, on one hand the interference from the D2D users fulfills the energy demands of the cellular users and on the other hand, this interference reduces the achievable rates of the cellular users. Hence, in order to guarantee the cellular users' demands, simultaneous information and power transmission is required to fulfill the network constraints. If the interference from the D2D users appear, improper Gaussian signaling helps in enlarging the achievable rate region. Figure \ref{BcChannelWithEH} illustrates the rate region improvement by allowing improper Gaussian signaling. Considering energy demands, the rate region of DPC is improved by improper Gaussian signaling which is depicted in Fig. \ref{fig:DPCwithImproperSig2}. When the BS utilizes DPC, it codes the transmit signal in a way that the received signal in one user is free from the interference from the other user. This type of coding is beneficial from information rate perspective but it is detrimental from the energy viewpoint. In this case, if the cellular users' energy demands are high enough, DPC becomes an inefficient coding scheme. The inefficiency of DPC is shown in Fig.~\ref{fig:DPCwithImproperSig2}, where the rate region of the cellular users is almost the same as the case of not utilizing DPC.
\subsection{Full-Duplex D2D}
\begin{figure*}
\centering
\begin{minipage}[b]{0.47\textwidth}
\subfigure[Rate-energy region of a full-duplex node in the network. Other users are neither demanding information nor energy.]{
\tikzset{every picture/.style={scale=.85}, every node/.style={scale=.8}}%
\begin{tikzpicture}

\begin{axis}[%
width=3.5in,
height=2.5in,
scale only axis,
xmin=0,
xmax=0.55,
xmajorgrids,
ymin=0,
ymax=1.56,
ymajorgrids,
xlabel={$R^{'}_1$ (bits/channel use)},
ylabel={$E^{'}_1$ (Energy unit)},
ylabel near ticks,
legend style={at={(axis cs: 0.55,1.56)},anchor= north east, draw=black,fill=white,legend cell align=left}
]
\addplot[color=blue,mark=o] plot table[row sep=crcr,]
{0	1.55057235503855\\
0.0783390551438543	1.35694432491807\\
0.118667632541724	1.16331722091386\\
0.143258017356836	0.969689309488715\\
0.173464368771613	0.656283376829296\\
0.218379407205781	0.37283297918268\\
0.278973338558668	0.190403925339105\\
0.349510993565266	0.0802399330987982\\
0.4005	0\\
};
\addlegendentry{PS, Proper sig. rate-energy region frontier};
\addplot[color=red,mark=*] plot table[row sep=crcr,]
{0	1.55057235503855\\
0.0817504514382899	1.35694432491807\\
0.132535669206069	1.16331722091386\\
0.175325055751165	0.969689309488715\\
0.219188528366315	0.776061389267602\\
0.269705840336762	0.582433840339989\\
0.332309502893398	0.388806127056778\\
0.41434311811609	0.195178300579788\\
0.528261247040493	0.00155057252809126\\
0.4005	0\\
};
\addlegendentry{PS, Improper sig. rate-energy region frontier};
\addplot[color=black,mark=asterisk] plot table[row sep=crcr,]
{0	1.55057235503855\\
0.4005	0\\
};
\addlegendentry{TS between EH and ID phases};

\end{axis}
\end{tikzpicture}%

\label{rateEnergy1}
}
\end{minipage}
\quad
\begin{minipage}[b]{0.47\textwidth}
\subfigure[Improvement in the rate-energy region for ID in one full-duplex node and EH in the other.]{
\tikzset{every picture/.style={scale=.85}, every node/.style={scale=.8}}%
\begin{tikzpicture}

\begin{axis}[%
width=3.5in,
height=2.5in,
scale only axis,
xmin=0,
xmax=0.4,
xmajorgrids,
ymin=0,
ymax=3,
ymajorgrids,
xlabel={$R^{'}_1$ (bits/channel use)},
ylabel={$E^{'}_2$ (Energy unit)},
ylabel near ticks,
legend style={at={(axis cs: 0,0)},anchor= south west, draw=black,fill=white,legend cell align=left}
]
]
\addplot [color=red,solid,mark=square]
  table[row sep=crcr]{
0	2.73146698940528\\
0.0136394614999997	2.73146698940528\\
0.26307741336562  	2.1046214345782\\
0.34358243524287 	1.20254733796034\\
0.362628265401238	0.725253617446479\\
0.373814270728985	0.46726731432495\\
0.37806820207561	    0.340697348204239\\
0.378067795697207	0\\
};
\addlegendentry{Proper sig. $R_1=R_2\geq0.2$};
\addplot [color=blue,dashed,mark=asterisk]
  table[row sep=crcr]{
0	2.73146698940528\\
0.0236679299534265	2.73146698940528\\
0.331635344633427	2.1046214345782\\
0.347915221263089	1.20254733796034\\
0.365422815914581	0.725253617446479\\
0.374598508003788	0.46726731432495\\
0.3781870460945	    0.340697348204239\\
0.378186639839401	0\\
};
\addlegendentry{Improper sig. $R_1=R_2\geq0.2$};
\end{axis}
\end{tikzpicture}%
\label{Fd_D2d_Tdma2}
}
\end{minipage}
\caption{Rate-energy region of the full-duplex nodes.}
\end{figure*}
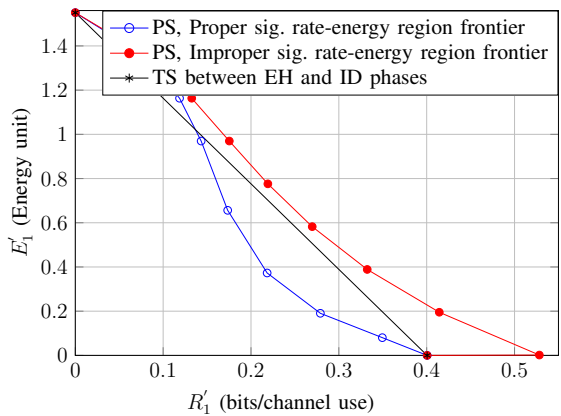
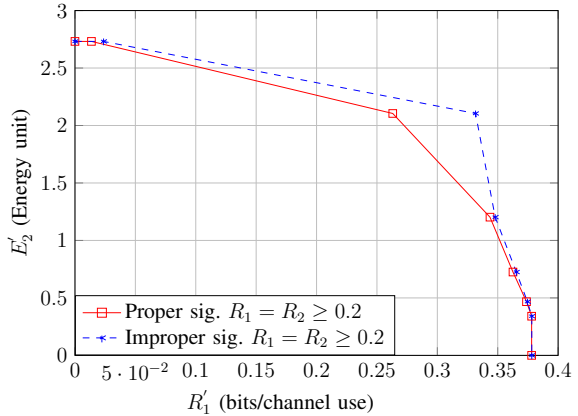

The performance of full-duplex D2D users is evaluated in this subsection. We consider the case, where D2D users behave as underlay cognitive radios. Hence, they are allowed to be active just in case that the demands of the primary users (cellular users) are fulfilled. Having this in mind that the primary users are supposed to fulfill certain rate constraints, D2D users maximize the achievable rates and energies. By utilizing improper signaling, the rate region of the D2D users is enlarged as shown in Fig. \ref{Fd_D2d_1}. Rate-energy region for full-duplex communication is studied, where PS and TS are the joint ID and EH techniques. Assuming PS receiver (refer to problem (\ref{B1})), in order to achieve maximum rate at a full-duplex user, BS needs to be silent. This is due to the fact that, maximum power delivered to the D2D user is achieved when the BS transmits with maximum power. Consider a case where a full-duplex D2D user runs out of power. Then, it is optimal to restrict the operating at the receivers to EH mode. Moreover, the transmitters need to direct their beams into the direction that delivers maximum power for that user (maximum ratio transmission in direction of the user). This operating point is depicted in Fig.~\ref{rateEnergy1}, where the plots cross the vertical axis. The maximum rate for a full-duplex node is achieved when the BS forms its beam so that the least power hits the user (achieved by zero-forcing). This can be observed in Fig.~\ref{rateEnergy1}, where the plots cross the horizontal axis. According to this figure, the efficiency of improper Gaussian signaling from the rate and energy perspective is vivid. If we share time (TS structure) between EH and ID phases in a particular D2D user, the black solid line consists of the outermost achievable rate and energy tuples. Moreover, we study the trade-off between the rates and energies of two full-duplex D2D users as well, where the first user is assumed to purely harvest energy and the second user requires information only and vice versa. The numerical results of the achievable rate-energy region for this strategy is depicted in Fig.~\ref{Fd_D2d_Tdma2}, (refer to problem (\ref{B112})). Intuitively, maximum ratio transmission (MRT) beamforming at the BS and D2D users toward the first D2D user, maximizes the received signal energy at the first D2D user, i.e., $E^{'}_1$, while this type of transmission is not rate optimal for the second D2D users, i.e., $R^{'}_2$. Thus, due to the priority weights of the rate and the energy optimization, all the points on the rate-energy region boundaries are achievable by optimum beamforming vectors.

\subsection{Joint rate and energy maximization}
In this subsection, we discus the performance of the investigated setup, when each receive antenna either decodes information or extracts the energy of the incident RF signal. 
Cellular users are able to harvest energy from the RF signal in the environment and decode information simultaneously with maximum power through AS receiver structure. However, D2D users are equipped with a single receive antenna and the receivers consume the whole received signal with its maximum power for ID purpose. In other words, D2D users are not demanding energy at a particular time and their main concern is information, (refer to problem (\ref{JointOpt})).\\
For two cellular users (demanding information and energy) and two D2D users (demanding information only), some interesting operating point on the boundary of six-dimensional rate-energy region is depicted in Table \ref{6dPareto}. According to this table, by allowing improper Gaussian signaling, the achievable sum rate can be improved compared to proper Gaussian signaling. This improvement in the achievable sum rate can be manifested in the energy as well. That means, considering the rate demands to be fulfilled by proper Gaussian signaling, the users can harvest more energy if the transmission scheme is improper Gaussian signaling.

\section{Multiple Users}
In this section we discuss the performance of a full-duplex interference channel (FD-IC, $J>2$) in proximity of a broadcast channel with multiple users $K>2$.
\subsection{Cellular Users}
In order for the BS to convey independent messages to downlink users in a single channel use, the number of transmit antennas at the BS should be at least as many as $K$, (i.e., $N=K$). Hence, due to limited power available at the BS, power allocation for the messages of the cellular users results in lower achievable rates for individual users. Thus, the $K$-dimensional rate region (e.g., Fig.~\ref{BcChannelNoEH} for $K=2$) shrinks as the number of users increase. This region tends to shrink as a function of EH demands and the channel realizations as well. That means, for high EH demands, the BS becomes unable of satisfying them, thus FD users get activated. The undesired interference imposed by FD users to satisfy the cellular users demand results in lower achievable rates for cellular users.
\subsection{Full-Duplex Users}
Presence of multiple FD communication pairs in proximity aids them in satisfying high EH demands on one hand. On the other hand, the individual achievable rates will diminish due to the undesired interference in signal detection and decoding. Assuming time-sharing for ID and EH purposes among any pair of FD users (at a particular time instant, one user does ID and the other user does EH), a 2D rate-energy region can be derived (e.g., Fig.~\ref{Fd_D2d_Tdma2}). As the number of FD users increase (compared to $J=2$ in Fig. \ref{Fd_D2d_Tdma2}), the intersection of the Pareto boundary with $R^{'}_2$-axis will remain unchanged. This is due to zero-forcing by the other users (i.e., transmit in the null space of the channel). Meanwhile the Pareto boundary will intersect the $E^{'}_1$-axis at higher energy level. This is due to an increase in the number of energy providers.    

\begin{table}[t!]
\centering
\begin{tabular}{|l|l|l|l|l|l|l|}
\hline
\multicolumn{7}{c}{Optimal Gaussian signalig}\\ \hline
Signaling & $r^{'}_1$ & $r^{'}_2$ & $r_1$ & $r_2$ & $e_1$ & $e_2$ \\  \hline
  Proper & 0.08 & 0.54 & 0.54 & 0.54 & 3.23 & 2.88 \\ \hline
  Improper & 0.11 & 0.61 & 0.71 & 0.61 & 3.23 & 2.88 \\ \hline
  \multicolumn{7}{c}{}\\ \hline
Signaling & $r^{'}_1$ & $r^{'}_2$ & $r_1$ & $r_2$ & $e_1$ & $e_2$ \\  \hline
  Proper & 0.00 & 0.43 & 0.72 & 0.54 & 2.52 & 2.56 \\ \hline
  Improper & 0.02 & 0.49 & 0.83 & 0.67 & 2.52 & 2.56 \\ \hline
\end{tabular}
\caption{Pareto-optimal operating points for different Chebyshev weights.}
\label{6dPareto}
\end{table}
\vspace*{-0.3cm}  
\section{Conclusion}
In this paper, we studied the rate and energy performance of a two-tier network which is composed of multiple full-duplex device-to-device communications incorporated in a macro-cell with a base station. Furthermore, we investigated the advantage of full-duplex D2D users in aiding the cellular users. Due to the energy and information demands of the users, different practical receiver structures for joint energy harvesting and information detection are investigated, namely, antenna separation, power splitting and time sharing. The performance of these types of receivers are manifested while improper Gaussian signaling is proposed to be utilized at the transmitters. we observed that, if the energy demands of the cellular users is out of the capability of the BS, the full-duplex D2D users aid in fulfilling the demands and it is not necessary to utilize non-linear dirty paper coding at the BS in case of high-enough energy demands. The achievable rate region of the users in the network and the achievable rate-energy region of the full-duplex users are studied and the optimal beamforming and resource allocation solutions are delivered.  

\section{Appendix} \label{App:Appendix}
For converting the problem into a SDP, we use the following Lemma.\\
{\bf Lemma 1},\cite{Guan2013}. The positive semi-definite constraint in (\ref{semiDefin1}) is satisfied if and only if ${\bf\hat{C}}_{x_j}=S_j{\bf\hat{t}}_{j}{\bf\hat{t}}_{j}^{T}\ \forall j\in\mathcal{D}$ and ${\bf\hat{C}}_{x_{B_k}}=S_{B_k}{\bf\hat{t}}_{B_k}{\bf\hat{t}}_{B_k}^{T}\ \forall k\in\mathcal{C}$. Notice that $S_j$ and $S_{B_k}$ are complex scalar variables satisfying $|S_j|\leq ||{\bf t}_j||^{2}$ and $|S_{B_k}|\leq ||{\bf t}_{B_k}||^{2}$, and ${\bf\hat{t}}_j=\frac{{\bf t}_j}{||{\bf t}_j||}$ and ${\bf\hat{t}}_{B_k}=\frac{{\bf t}_{b_k}}{||{\bf t}_{b_k}||}$, where ${\bf t}_j$ and ${\bf t}_{B_k}$ are defined as is \eqref{BfVectors}.\\
{\bf proof:} The proof can be found in~\cite{Guan2013}.\\
By using this lemma, optimizing over positive semi-definite matrices of sizes $M\times M$ and $N\times N$ reduces to optimizing over complex scalars, i.e.,  $S_j,\forall j$ and $S_{B_k},\forall k$. For convenience in formulation and readability, we consider two full-duplex users $J=2$, while formulation can be generalized for any arbitrary $J$. We rewrite the pseudo-variance of the received signal as,
{\small
\begin{align}
\hat C_{y_k}=&
\sum_{m=1}^{K}({\bf h}^{H}_{kB}{\bf\hat{t}}_{B_m})^{2}S_{B_m}+\sum_{j=1}^{2}({\bf h}^{H}_{kj}{\bf\hat{t}}_j)^{2}S_j,\ \forall k\in\mathcal{C},\\
\hat C_{z_j}=&
({\bf g}^{H}_{ji}{\bf\hat{t}}_i)^{2}S_i+ \sum_{m=1}^{K}({\bf g}^{H}_{jB}{\bf\hat{t}}_{B_m})^{2}S_{B_m}+\kappa({\bf g}^{H}_{jj}{\bf\hat{t}}_j)^{2}S_j, \ \forall j\in\mathcal{D},
\end{align}}
The pseudo-variance of the interference-plus-transmitter noise ($\hat C_{w_k}$ and $\hat C_{q_j}$) is written as,
{\small
\begin{align}
\hat C_{w_k}&=
\sum_{\substack{m=1\\ m\neq k}}^{K}({\bf h}^{H}_{kB}{\bf\hat{t}}_{B_m})^{2}S_{B_m} +\sum_{j=1}^{2}({\bf h}^{H}_{kj}{\bf\hat{t}}_j)^{2}S_j,\ \forall k\in\mathcal{C},\\
\hat C_{q_j}&=
\sum_{m=1}^{K}({\bf g}^{H}_{jB}{\bf\hat{t}}_{B_m})^{2}S_{B_m}+\kappa({\bf g}^{H}_{jj}{\bf\hat{t}}_j)^{2}S_j, \ \forall j\in\mathcal{D}.
\end{align}}
For simplicity in formulation and without loss of generality, we assume two active cellular users, i.e., $K=2$. We define the following vectors,
{\small
\begin{align}
&{\bf s}=[S_1 \quad S_2 \quad S_{B_1} \quad S_{B_2}]^{T},\\
&{\bf a}_1=
C^{*^{-1}}_{y_1}\left[({\bf h}^{H}_{11}{\bf\hat{t}}_1)^{2}\quad ({\bf h}^{H}_{12}{\bf\hat{t}}_2)^{2}\quad ({\bf h}^{H}_{1B}{\bf\hat{t}}_{B_1})^{2}\quad ({\bf h}^{H}_{1B}{\bf\hat{t}}_{B_2})^{2}\right]^{H},\\
&{\bf a}_2=
C^{*^{-1}}_{y_2}\left[({\bf h}^{H}_{21}{\bf\hat{t}}_1)^{2}\ ({\bf h}^{H}_{22}{\bf\hat{t}}_2)^{2}\ ({\bf h}^{H}_{2B}{\bf\hat{t}}_{B_1})^{2}\ ({\bf h}^{H}_{2B}{\bf\hat{t}}_{B_2})^{2}\right]^{H},\\
&{\bf a}^{'}_1=
C^{*^{-1}}_{z_1}\left[\kappa({\bf g}^{H}_{11}{\bf\hat{t}}_1)^{2}\ ({\bf g}^{H}_{12}{\bf\hat{t}}_2)^{2}\ ({\bf g}^{H}_{1B}{\bf\hat{t}}_{B_1})^{2}\ ({\bf g}^{H}_{1B}{\bf\hat{t}}_{B_2})^{2}\right]^{H},\\
&{\bf a}^{'}_2=
C^{*^{-1}}_{z_2}\left[({\bf g}^{H}_{21}{\bf\hat{t}}_1)^{2}\ \kappa({\bf g}^{H}_{22}{\bf\hat{t}}_2)^{2}\ ({\bf g}^{H}_{2B}{\bf\hat{t}}_{B_1})^{2}\ ({\bf g}^{H}_{2B}{\bf\hat{t}}_{B_2})^{2}\right]^{H}.
\end{align}}
We also define the transmit noise covariance matrices and the corresponding interference vectors as,
{\small
\begin{align}
&{\bf b}_1=
C^{*^{-1}}_{y_1}\left[({\bf h}^{H}_{11}{\bf\hat{t}}_1)^{2}\quad ({\bf h}^{H}_{12}{\bf\hat{t}}_2)^{2}\quad 0\quad ({\bf h}^{H}_{1B}{\bf\hat{t}}_{B_2})^{2}\right]^{H},\\
&{\bf b}_2=
C^{*^{-1}}_{y_2}\left[({\bf h}^{H}_{21}{\bf\hat{t}}_1)^{2}\quad ({\bf h}^{H}_{22}{\bf\hat{t}}_2)^{2}\quad ({\bf h}^{H}_{2B}{\bf\hat{t}}_{B_1})^{2}\quad 0\right]^{H},\\
&{\bf b}^{'}_1=
C^{*^{-1}}_{z_1}\left[\kappa({\bf g}^{H}_{11}{\bf\hat{t}}_1)^{2}\quad 0\quad ({\bf g}^{H}_{1B}{\bf\hat{t}}_{B_1})^{2}\quad ({\bf g}^{H}_{1B}{\bf\hat{t}}_{B_2})^{2}\right]^{H},\\
&{\bf b}^{'}_2=
C^{*^{-1}}_{z_2}\left[0\quad \kappa({\bf g}^{H}_{22}{\bf\hat{t}}_2)^{2}\quad ({\bf g}^{H}_{2B}{\bf\hat{t}}_{B_1})^{2}\quad ({\bf g}^{H}_{2B}{\bf\hat{t}}_{B_2})^{2}\right]^{H}.
\end{align}}
We define the matrices ${\bf A}$, ${\bf A}^{'}$, ${\bf B}$, ${\bf B}^{'}$,  and ${\bf S}$ as,
{\small
\begin{align}
{\bf A}_k= {\bf a}_k{\bf a}^{H}_k, \qquad {\bf A}^{'}_j= {\bf a}^{'}_j{\bf a}^{'^{H}}_j,\quad
{\bf B}_k= {\bf b}_k{\bf b}^{H}_k, \qquad {\bf B}^{'}_j= {\bf b}^{'}_j{\bf b}^{'^{H}}_j,\quad
{\bf S}= {\bf s}{\bf s}^{H}.
\end{align}}
By the defined vectors and matrices, we can state the following equalities,
{\small
\begin{align}
C^{-2}_{y_k}|{\hat{C}}_{y_k}|^{2}=|{\bf a}^{H}_k{\bf s}|^{2}={\rm{Tr}}({\bf A}_k \bf S),\quad
C^{-2}_{w_k}|{\hat{C}}_{w_k}|^{2}=|{\bf b}^{H}_k{\bf s}|^{2}={\rm{Tr}}({\bf B}_k \bf S),\\
C^{-2}_{z_j}|{\hat{C}}_{z_j}|^{2}=|{\bf a}^{'^{H}}_j{\bf s}|^{2}={\rm{Tr}}({\bf A}^{'}_j \bf S),\quad 
C^{-2}_{q_j}|{\hat{C}}_{q_j}|^{2}=|{\bf b}^{'^{H}}_j{\bf s}|^{2}={\rm{Tr}}({\bf B}^{'}_j \bf S).
\end{align}}
Considering lemma 1 and aforementioned equalities, we reformulate (\ref{r3}) as,
{\small
\begin{align}
\Gamma=\Lambda-\lambda_{\star}&\leq \frac{1}{2\alpha_i}\log\left(\frac{1-{\rm{Tr}}({\bf A}_k{\bf S})}{1-{\rm{Tr}}({\bf B}_k{\bf S})}\right).
\end{align}}
Constraints (\ref{Pseudo39B}) can also be reformulated as,
{\small
\begin{align}
{\rm {Tr}}({\bf M}_j{\bf S})\leq ||{\bf t}_j||^{4},\ \forall j\in\{1,2\},\quad
{\rm {Tr}}({\bf M}_k{\bf S})\leq ||{\bf t}_{B_k}||^{4},\ \forall k\in\{1,2\}
\end{align}}
where ${\bf M}_i={\bf m}_i{\bf m}^{T}_i$ and ${\bf m}_i$ is the $i^{th}$ column of $4\times 4$ identity matrix.\\
Therefore, the optimization problem (\ref{A6}) can be written as a SDP as follows:
{\small
\begin{subequations}\label{A77}
\begin{align}
\hspace{-.1cm}\max_{\Gamma, {\bf S}\succeq \bf 0}\quad  &\Gamma \tag{\ref{A77}}\\
{\rm{s.t.}}\quad &\Gamma\leq \frac{1}{2\alpha_k}\log\left(\frac{1-{\rm {Tr}}({\bf A}_k{\bf S})}{1-{\rm {Tr}}({\bf B}_k{\bf S})}\right),\ \forall k\in\{1,2\},\label{L1}\\
&{\rm {Tr}}({\bf M}_k{\bf S})\leq ||{\bf t}_{B_k}||^{4},\ \forall k\in\{1,2\},\label{sdp222}\\
&{\rm {Tr}}({\bf M}_j{\bf S})\leq ||{\bf t}_j||^{4},\ \forall j\in\{1,2\},\label{sdp2}
\end{align}
\end{subequations}}
where the rank-1 constraint of $\bf S$ is dropped. Thereof the solution is an upper bound for the original problem, unless the optimal $\bf S$ is intrinsically rank-1. If optimal $\bf S$ has a higher rank, a sub-optimal rank-1 solution can be obtained by the Gaussian randomization procedure~\cite{Jorswieck2011,Poor2011}.\\
{\bf Theorem 1},\cite{Guan2013}. {\it For any matrix $\bf S$ that satisfies (\ref{sdp2}), the following inequalities fulfill},
{\small
\begin{align}
&1-{\rm{Tr}}({\bf A}_k{\bf S})\geq C^{-2}_{y_k}\sigma^{4}\geq 0,\ \forall k\in\{1,2\},\label{T1}\\
&1-{\rm{Tr}}({\bf B}_k{\bf S})\geq C^{-2}_{w_k}\sigma^{4}\geq 0,\ \forall k\in\{1,2\}\label{T2}.
\end{align}}
If (\ref{T1}) and (\ref{T2}) fulfills, then problem (\ref{A6}) becomes a quasi-convex problem and can be solved by bisection \cite{Boyd2004}. We consider the following feasibility problem by bisecting over $\Gamma$.
{\small
\begin{align}
&\hspace{-.6cm}{\rm {find}}\quad {\bf S}\in\mathbb{S}^{2}\quad \rm{s.t.}\ (\ref{L1})-(\ref{T2}),\label{A8}
\end{align}}
where ${\bf S}^{\star}$ is found with a certain bisection accuracy. If the solution, i.e. ${\bf S}^{\star}$ is rank-1 the ${\bf s}^{\star}$ can be calculated by eigen-value decomposition. Then, by replacing the elements of ${\bf s}^{\star}$, i.e., $S_j$ and $S_{B_k},\ \forall j,k\in\{1,2\}$, in the equation of lemma 1, that is, ${\bf\hat{C}}_{x_j}=S_j{\bf\hat{t}}_{j}{\bf\hat{t}}_{j}^{T},\ \forall j\in\{1,2\}$ and ${\bf\hat{C}}_{x_{B_k}}=S_{B_k}{\bf\hat{t}}_{B_k}{\bf\hat{t}}_{B_k}^{T},\ \forall k\in\{1,2\}$, the optimal pseudo-covariance matrices are delivered.
\bibliographystyle{IEEEtran} 
\bibliography{reference}

\begin{thebibliography}{10}
\providecommand{\url}[1]{#1}
\csname url@samestyle\endcsname
\providecommand{\newblock}{\relax}
\providecommand{\bibinfo}[2]{#2}
\providecommand{\BIBentrySTDinterwordspacing}{\spaceskip=0pt\relax}
\providecommand{\BIBentryALTinterwordstretchfactor}{4}
\providecommand{\BIBentryALTinterwordspacing}{\spaceskip=\fontdimen2\font plus
\BIBentryALTinterwordstretchfactor\fontdimen3\font minus
  \fontdimen4\font\relax}
\providecommand{\BIBforeignlanguage}[2]{{%
\expandafter\ifx\csname l@#1\endcsname\relax
\typeout{** WARNING: IEEEtran.bst: No hyphenation pattern has been}%
\typeout{** loaded for the language `#1'. Using the pattern for}%
\typeout{** the default language instead.}%
\else
\language=\csname l@#1\endcsname
\fi
#2}}
\providecommand{\BIBdecl}{\relax}
\BIBdecl

\bibitem{Bliss2014}
A.~Sabharwal, P.~Schniter, D.~Guo, D.~Bliss, S.~Rangarajan, and R.~Wichman,
  ``{In-Band Full-Duplex Wireless: Challenges and Opportunities},'' \emph{IEEE
  J. Sel. Areas Commun.}, vol.~32, no.~9, pp. 1637--1652, Sept 2014.

\bibitem{Sabharwal2014}
E.~Everett, A.~Sahai, and A.~Sabharwal, ``{Passive Self-Interference
  Suppression for Full-Duplex Infrastructure Nodes},'' \emph{IEEE Trans.
  Wireless Commun.}, vol.~13, no.~2, pp. 680--694, February 2014.

\bibitem{Shankar2012}
M.~Duarte, A.~Sabharwal, V.~Aggarwal, R.~Jana, K.~Ramakrishnan, C.~Rice, and
  N.~Shankaranarayanan, ``{Design and Characterization of a Full-Duplex
  Multiantenna System for WiFi Networks},'' \emph{IEEE Trans. Veh. Tech.},
  vol.~63, no.~3, pp. 1160--1177, March 2014.

\bibitem{Shankar20122}
A.~Sahai, G.~Patel, C.~Dick, and A.~Sabharwal, ``{Understanding the impact of
  phase noise on active cancellation in wireless full-duplex},'' in
  \emph{Conference Record of the Forty Sixth Asilomar Conf. on Signals, Systems
  and Computers}, Nov 2012, pp. 29--33.

\bibitem{Bliss2012}
B.~Day, A.~Margetts, D.~Bliss, and P.~Schniter, ``{Full-Duplex MIMO Relaying:
  Achievable Rates Under Limited Dynamic Range},'' \emph{IEEE J. Sel. Areas
  Commun.}, vol.~30, no.~8, pp. 1541--1553, September 2012.

\bibitem{Eltawil2015}
A.~Elsayed and A.~Eltawil, ``{All-Digital Self-Interference Cancellation
  Technique for Full-Duplex Systems},'' \emph{IEEE Trans. Wireless Commun.},
  vol.~14, no.~7, pp. 3519--3532, July 2015.

\bibitem{Hedley2008}
H.~Suzuki, T.~V.~A. Tran, I.~Collings, G.~Daniels, and M.~Hedley,
  ``{Transmitter Noise Effect on the Performance of a MIMO-OFDM Hardware
  Implementation Achieving Improved Coverage},'' \emph{IEEE J. Sel. Areas
  Commun.}, vol.~26, no.~6, pp. 867--876, August 2008.

\bibitem{Wichman2013}
M.~Vehkapera, T.~Riihonen, and R.~Wichman, ``{Asymptotic analysis of
  full-duplex bidirectional {MIMO} link with transmitter noise},'' in
  \emph{Proc. IEEE International Symposium on Personal Indoor and Mobile Radio
  Commun.}, Sept 2013, pp. 1265--1270.

\bibitem{Wang2014}
D.~Niyato and P.~Wang, ``{Delay-Limited Communications of Mobile Node With
  Wireless Energy Harvesting: Performance Analysis and Optimization},''
  \emph{IEEE Trans. Veh. Tech.}, vol.~63, no.~4, pp. 1870--1885, May 2014.

\bibitem{Ho2011}
R.~Zhang and C.~K. Ho, ``{MIMO Broadcasting for Simultaneous Wireless
  Information and Power Transfer},'' in \emph{Proc. IEEE Global Telecommun.
  Conf.}, Dec 2011, pp. 1--5.

\bibitem{Gunduz2013}
R.~Gangula, D.~Gesbert, and D.~Gunduz, ``{Optimizing feedback in energy
  harvesting {MISO} communication channels},'' in \emph{Proc. IEEE Global Conf.
  on Signal and Inf. Proc.}, Dec 2013, pp. 359--362.

\bibitem{ZhangMay2013}
R.~Zhang and C.~K. Ho, ``{MIMO Broadcasting for Simultaneous Wireless
  Information and Power Transfer},'' \emph{IEEE Trans. Wireless Commun.},
  vol.~12, no.~5, pp. 1989--2001, May 2013.

\bibitem{ZhouNov2013}
X.~Zhou, R.~Zhang, and C.~K. Ho, ``{Wireless information and power transfer:
  architecture design and rate-energy tradeoff},'' \emph{IEEE Trans. Commun.},
  vol.~61, no.~11, pp. 4754--4767, 2013.

\bibitem{BiApril2015}
S.~Bi, C.~K. Ho, and R.~Zhang, ``{Wireless powered communication: opportunities
  and challenges},'' \emph{IEEE Communications Magazine}, vol.~53, no.~4, pp.
  117--125, 2015.

\bibitem{Liu2013}
L.~Liu, R.~Zhang, and K.~C. Chua, ``{Wireless Information and Power Transfer: A
  Dynamic Power Splitting Approach},'' \emph{IEEE Transactions on
  Communications}, vol.~61, no.~9, pp. 3990--4001, September 2013.

\bibitem{Kariminezhad2017SPL}
A.~Kariminezhad, S.~Gherekhloo, and A.~Sezgin, ``Optimal power splitting for
  simultaneous information detection and energy harvesting,'' \emph{IEEE Signal
  Processing Letters}, vol.~24, no.~7, pp. 963--967, July 2017.

\bibitem{Jafar2009}
V.~Cadambe, S.~A. Jafar, and C.~Wang, ``{Interference Alignment with Asymmetric
  Complex Signaling-Settling the Host-Madson Nosratinia Conjecture},''
  \emph{IEEE Trans. Inf. Theory}, 2010.

\bibitem{Jorswieck2012}
Z.~Ho and E.~Jorswieck, ``{Improper Gaussian Signaling on the Two-User {SISO}
  Interference Channel},'' \emph{IEEE Trans. Wireless Commun.}, vol.~11, no.~9,
  pp. 3194--3203, September 2012.

\bibitem{Zhang2013}
Y.~Zeng, C.~Yetis, E.~Gunawan, Y.~L. Guan, and R.~Zhang, ``{Transmit
  Optimization With Improper Gaussian Signaling for Interference Channels},''
  \emph{EEE Trans. Signal Proc.}, vol.~61, no.~11, pp. 2899--2913, June 2013.

\bibitem{Guan2013}
Y.~Zeng, R.~Zhang, E.~Gunawan, and Y.~L. Guan, ``{Optimized Transmission with
  Improper Gaussian Signaling in the K-User MISO Interference Channel},''
  \emph{IEEE Trans. Wireless Commun.}, vol.~12, no.~12, pp. 6303--6313,
  December 2013.

\bibitem{Kariminezhad2017A}
A.~Kariminezhad, A.~Sezgin, and M.~Pesavento, ``{Power efficiency of improper
  signaling in MIMO full-duplex relaying for K-user interference networks},''
  in \emph{2017 IEEE International Conference on Communications (ICC)}, May
  2017, pp. 1--6.

\bibitem{Ottersten2014}
E.~Bjornson, E.~Jorswieck, M.~Debbah, and B.~Ottersten, ``{Multiobjective
  Signal Processing Optimization: The way to balance conflicting metrics in 5G
  systems},'' \emph{IEEE Signal Proc. Magazine}, vol.~31, no.~6, pp. 14--23,
  Nov 2014.

\bibitem{Luo2010}
Z.-Q. Luo, W.-K. Ma, A.-C. So, Y.~Ye, and S.~Zhang, ``{Semidefinite Relaxation
  of Quadratic Optimization Problems},'' \emph{IEEE Signal Proc. Magazine},
  vol.~27, no.~3, pp. 20--34, May 2010.

\bibitem{Quan2006}
Z.-Q. Luo and W.~Yu, ``{An introduction to convex optimization for
  communications and signal processing},'' \emph{IEEE J. Sel. Areas Commun.},
  vol.~24, no.~8, pp. 1426--1438, Aug 2006.

\bibitem{Sidi2006}
N.~Sidiropoulos, T.~Davidson, and Z.-Q. Luo, ``{Transmit beamforming for
  physical-layer multicasting},'' \emph{IEEE Trans. Signal Proc.}, vol.~54,
  no.~6, pp. 2239--2251, June 2006.

\bibitem{Gorokhov2004}
A.~Gorokhov, M.~Collados, D.~Gore, and A.~Paulraj, ``{Transmit/receive MIMO
  antenna subset selection},'' in \emph{Proc. IEEE International Conf. on
  Acoustics, Speech, and Signal Proc.}, vol.~2, May 2004, pp. ii--13--16 vol.2.

\bibitem{Gore2002}
D.~Gore and A.~Paulraj, ``{MIMO antenna subset selection with space-time
  coding},'' \emph{IEEE Trans. Signal Proc.}, vol.~50, no.~10, pp. 2580--2588,
  Oct 2002.

\bibitem{Schreier2010}
P.~Schreier and L.~Scharf, \emph{{Statistical Signal Processing of
  Complex-Valued Data: The Theory of Improper and Noncircular Signals}}.\hskip
  1em plus 0.5em minus 0.4em\relax Cambridge University Press, 2010.

\bibitem{Adali2011}
T.~Adali, P.~Schreier, and L.~Scharf, ``{Complex-Valued Signal Processing: The
  Proper Way to Deal With Impropriety},'' \emph{IEEE Trans. Signal Proc.},
  vol.~59, no.~11, pp. 5101--5125, Nov 2011.

\bibitem{Cover1991}
T.~M. Cover and J.~A. Thomas, \emph{Elements of Information Theory}.\hskip 1em
  plus 0.5em minus 0.4em\relax New York, NY, USA: Wiley-Interscience, 1991.

\bibitem{Boyd2004}
S.~Boyd and L.~Vandenberghe, \emph{{Convex Optimization}}.\hskip 1em plus 0.5em
  minus 0.4em\relax Cambridge University Press, 2004.

\bibitem{Vishwanath2003}
S.~Vishwanath, N.~Jindal, and A.~Goldsmith, ``{Duality, achievable rates, and
  sum-rate capacity of Gaussian MIMO broadcast channels},'' \emph{IEEE Trans.
  Inf. Theory}, vol.~49, no.~10, pp. 2658--2668, Oct 2003.

\bibitem{Vishwanath20032}
P.~Viswanath and D.~Tse, ``{Sum capacity of the vector Gaussian broadcast
  channel and uplink-downlink duality},'' \emph{IEEE Trans. Inf. Theory},
  vol.~49, no.~8, pp. 1912--1921, Aug 2003.

\bibitem{Jorswieck2011}
R.~Mochaourab and E.~Jorswieck, ``{Optimal Beamforming in Interference Networks
  with Perfect Local Channel Information},'' \emph{IEEE Trans. Signal Proc.},
  vol.~59, no.~3, pp. 1128--1141, March 2011.

\bibitem{Poor2011}
X.~C. Shang and V.~B.~Poor, ``{Multiuser {MISO} Interference Channels With
  Single-User Detection: Optimality of Beamforming and the Achievable Rate
  Region},'' \emph{IEEE Trans. Inf. Theory}, vol.~57, no.~7, July 2011.

\end{thebibliography}
\end{document}